\documentclass[9pt,twocolumn,twoside,final]{pnas-new-nolinenum}

\articletype{PHYSICS}

\templatetype{pnasresearcharticle}

\begin{document}

\title{Overlap locking and non-perturbative effects in spin glasses}

\author[a,b,c]{Silvio~Franz}
\author[c,d,1]{Giorgio~Parisi}
\author[d,c]{Federico~Ricci-Tersenghi}

\affil[a]{Dipartimento di Matematica e Fisica ``Ennio De Giorgi'', Università del Salento, and INFN Sez. Lecce, Via per Arnesano, Lecce, 73100, Italy}
\affil[b]{Université Paris-Saclay, CNRS, LPTMS, 91405 Orsay, France}
\affil[c]{International Research Center of Complexity Sciences, Hangzhou International Innovation Institute, Beihang University, Hangzhou 311115, China}
\affil[d]{Dipartimento di Fisica, Sapienza Universit\`a di Roma, CNR--Nanotec, Rome Unit, and INFN--Sezione di Roma 1, 00185 Rome, Italy}


\significancestatement{The replica method has allowed the construction of a coherent mean-field theory for spin glasses. However, a large number of fundamental problems, both perturbative and non-perturbative, remain to be addressed to build a theory of fluctuations in finite-dimensional systems. Here, we take a step toward understanding non-perturbative effects: we settle the problem of finite-size corrections in the Sherrington-Kirkpatrick spin glass and compute the exponents for the decay of correlations in Dyson's hierarchical model of spin glasses. The values that we find are in excellent agreement with numerical simulations. The crucial step is to control what happens to the order parameter when two otherwise independent systems are coupled via an interaction that grows as $N^\alpha$ with $0<\alpha<1$.
}

\authordeclaration{The authors declare no competing interest.}
\equalauthors{All authors contributed equally to this work}
\correspondingauthor{\textsuperscript{1}To whom correspondence should be addressed. E-mail:  giorgio.parisi@gmail.com}


\begin{abstract}
We study the phenomenon of the locking of the order parameter (or synchronization) in spin glasses at low temperatures. When two systems with independent disorders are coupled, their overlaps become similar. A crucial question is how this effect depends on the strength of the coupling between the two systems. Non-perturbative phenomena are present when $1 \ll \Delta H \ll N$, being $\Delta H$ the coupling Hamiltonian and $N$ the system size.
In this intermediate-coupling region, the effect is related to finite-size free-energy corrections in mean-field spin-glass models and to correlations in the Dyson hierarchical spin glass, a model that mimics the physics of finite-dimensional systems. 
We study this phenomenon in the mean-field approach, both analytically and numerically, and we finally compute the critical exponents for finite-volume corrections in mean-field theory and for the decay of correlations in the Dyson hierarchical model.
\end{abstract}

\dates{This manuscript was compiled on \today}
\doi{\url{www.pnas.org/cgi/doi/10.1073/pnas.XXXXXXXXXX}}

\maketitle

\thispagestyle{firststyle}
\ifthenelse{\boolean{shortarticle}}{\ifthenelse{\boolean{singlecolumn}}{\abscontentformatted}{\abscontent}}{}

\Firstpage

Let us consider a straightforward question in spin glasses. We are in the spin-glass phase at low temperatures. We take two samples in dimension $D$, each of size $L^D$, and glue them together to form a new sample of size $2L$ in one direction and $L$ in the other $D-1$ directions. What are the typical correlations among spins in the two interacting samples? Do they decay exponentially with $L$, or do they have a power law decay, and what is the decay exponent?

Straightforward question, but finding the answer is tough. It amounts to computing the decay of the correlation functions in the low-temperature phase.

We can simplify the question. We consider two mean-field models with $N=L^D$ spins, and couple the two systems via a Hamiltonian $\Delta H$ of order $L^{D-1}=N^{(D-1)/D}\equiv N^\alpha$ with $\alpha=\frac{D-1}{D}$. This \textit{contact} Hamiltonian is much larger than one, but it is sub-extensive, $1 \ll \Delta H \ll N$.

Finding the answer remains difficult. When $\Delta H=O(1)$, we can use a perturbative expansion. By extrapolating, we may obtain information about what happens near $\alpha=0$. When $\Delta H=O(\epsilon N)$, the Hamiltonian is extensive: we could get information on the behavior near $\epsilon=0$. This paper aims to answer this question by combining analyses from the two different limiting regions. This allows us to present a non-perturbative computation of the critical exponents in the low temperature phase. The computation is not exact for short-range models (though it may be exact for long-range ones); however, it is an important step forward in understanding non-perturbative effects.

Here, we focus on the low-temperature phase of spin glasses (and, more generally, disordered systems) and study the problem within the replica approach. It is well known that, in the replica approach to spin glasses and in numerical simulations, long-range correlations arise from the spontaneous breaking of a continuous symmetry. From an abstract viewpoint, the situation is similar to a standard ferromagnetic model with $O(n)$ symmetry. In both cases, the symmetry group is continuous and broken at low temperature. Unfortunately, there are crucial differences:

    In the ferromagnetic case, at low temperatures, we have a spontaneous magnetization $m_a$ ($a=1,\ldots,n$). The unbroken residual symmetry group is $O(n-1)$, and we have $n-1$ Goldstone bosons with a correlation function in momentum space proportional to $1/k^2$. 
    
    \Endparasplit
    The interaction of these Goldstone bosons is described by the non-linear sigma model, where one introduces fields $\sigma_a$ with fixed norm (i.e., $\sum_{a=1,n}\sigma_a^2=1$).
    Using Ward identities, one can show that the interaction of these gapless modes does not produce infrared divergences in space dimensions $D$ greater than $2$, and therefore the low-temperature phase is well-defined. Moreover, in dimensions $D=2+\epsilon$, we can construct an $\epsilon$ expansion for the critical exponents \cite{parisi:88,zinn2021quantum}.
    
    In spin glasses, the symmetry group is the permutation group of $n$ elements $S(n)$, which in the replica limit $n\to 0$ acquires characteristics of a continuous group. The order parameter is an $n \times n $ matrix $Q$. The residual unbroken symmetry group is a mess, and we do not know how to parameterize it. There is an infinite number of Goldstone bosons whose most singular correlation functions are
    proportional to $1/k^4$ \cite{de1984spin,temesvari1988long}. A full one-loop computation may involve millions of terms, and it has never been done. 
    Last, but not least, non-perturbative phenomena are not well understood either, although an interface tension computation suggests that the lower critical dimension is $D_\text{lcd}=5/2$ \cite{franz1994interfaces,astuti2019new}.

Here, we will show how to perform the most straightforward non-perturbative computation for spin glasses and apply it to physical problems. We will compute a basic integral, and we will show that the knowledge of this integral can be used to solve these two long-standing problems:
\begin{itemize}
    \item The finite-size corrections to the internal energy of the Sherrington-Kirkpatrick model. These have been studied numerically. The model exhibits power-law corrections, but the exponent could not be predicted from first principles.
    \item The power-law decay of the correlation function with distance in the Dyson-like hierarchical model for spin glasses \cite{franz2009overlap}. This is particularly relevant because the hierarchical model is the simplest one for which the renormalization group can be applied: indeed, the first renormalization group computations of the critical exponents were done by Wilson precisely in the hierarchical model \cite{wilson:74}, which he rediscovered, the relation among the dimension $D$ of the the space and the hierarchical model is already in Wilson's seminal papers. Only later, in \cite{bleher1975critical}, Bleher and Sinai realized the deep relation between Wilson's real-space renormalization group and the hierarchical model. On a different note, the hierarchical model should have exponents similar to those of the one-dimensional model with long-range interaction. Also, this model has been extensively studied as the best proxy for $D$-dimensional systems \cite{katzgraber:03, leuzzi:08, leuzzi:09, jensen:21}. Some of the qualitative features we find may thus apply to short-range models.
\end{itemize}

\section{The mode locking integral}

Let us start with a propaedeutic discussion of the ferromagnetic $O(3)$ model. In the low-temperature region, the magnetization is $m\,\vec{n}$, where $ \vec {n} $ is a unit vector on the sphere. We now take two coupled systems  ($\sigma^L$ and $\sigma^R$), for simplicity of equal size $N$, and introduce an interaction term $\frac{z}{N^2}\left( \sum_{i=1}^N \vec\sigma_i^L -\sum_{i=1}^N \vec\sigma_i^R\right)^2$.
The free energy of the compound system reads
\begin{eqnarray}
    && {\cal F}(z,N) = -\log\left({\cal Z}(z,N)/Z^2(N)\right) = -2Nf \label{eq:prima} \\
    && -\log\left(\int d\vec{m}^L\,d\vec{m}^R \; e^{ -N (f(\vec{m}^L)+ f(\vec{m}^R))-z |\vec{m}^L-\vec{m}^R|^2}\right) \nonumber
\end{eqnarray}
where the temperature is included in the definition of $f$ to lighten the notation.
Indeed, the function $f(\vec{m})$ is the free energy as a function of the order parameter, which has a Mexican hat shape with a minimum for $|\vec{m}|= m_*$. 
If $z$ is independent of $N$, in the limit $N\to\infty$ we get
\begin{multline}
    \lim_{N\to\infty} {\cal Z}(z,N)/Z^2(N) \equiv {\cal Z}(z) =
    4\pi \int d\vec{n}\;e^{-z m_*^2 (\vec{n}_0-\vec{n})^2}\\
    = 4\pi\int dg\; e^{-z m_*^2 (\vec{n}_0-\vec{n}(g))^2}\,,
    \label{eq:O3}
\end{multline}
where $\vec{n}_0$ is an arbitrary vector on the unit sphere. We call ${\cal Z}(z,N)$ the mode locking integral for the model.
We could write Eq.~\ref{eq:O3} as an integral over the possible rotation $g$ of the $O(3)$ group with $\vec{n}(g)=R^{-1}(g) \vec{n}_0 R(g)$. The rotation group structure allows us to do the computation and get
\begin{multline}
{\cal Z}(z) = 8\pi^2 \int_0^{\pi}\sin(\theta) d\theta\; e^{-2z m_*^2(1-\cos(\theta))}=\\
= 4\pi^2\; e^{-2zm_*^2}\;\frac{\sinh(2zm_*^2)}{zm_*^2}\,.
\label{eq:Z}
\end{multline}
The result is simple, and it can be generalized to the $O(n)$ symmetry without difficulty.

We can also compute quite straightforwardly the mode locking integral when $z=tN$ with $t$ finite, and the value of the two magnetizations at the saddle point is equal.
Denoting by ${\cal H}$ the Hessian of small fluctuations, the $t$-dependent part of ${\cal F}$ can be written as 
\begin{eqnarray}
    {\cal F} (z=tN,N)=-\frac{1}{2}\text{Tr}\log({\cal H}+2t\,\mathbb{I})\,.
    \label{eq:hess}
\end{eqnarray}

A central quantity in our computation is 
\begin{equation}
{\cal D}(z)=\langle |
\vec{m}^L-\vec{m}^R|^2\rangle=\frac{d\log {\cal Z}(z)}{dz}=\frac{d{\cal F}(z)}{dz}\,.
\end{equation}
This quantity, as discussed in the Supplementary Information, is a proxy for the \emph{connected correlation function} between the two subsystems. 
In this model, this quantity is also related to the spectrum of the Hessian $\mathcal{H}$.
From Eq.~\ref{eq:Z} we find that ${\cal D}(z)\sim 1/z$, for large $z$, implying the two magnetizations $\vec{m}^L$ and $\vec{m}^R$ to become equal, which is called locking (or synchronization) of the two systems.
The same behavior holds in the limit $z=tN\to\infty$ with $t=O(1)$, which can be obtained from Eq.~\ref{eq:hess} as a consequence of the zero modes. As we will see, such a behavior is associated (neglecting logs) with $O(N^{-1})$ finite-size corrections to the free-energy. 

When we try to do the equivalent computation in spin glasses, the order parameter is an  $n\times n$ matrix.
At the end, we face the following integral in the $n\to 0$ limit (apart from $z$-independent terms similar to those in Eq.~\ref{eq:prima})
\begin{multline}
    \hspace{-4mm}
    {\cal F}(z,N)\equiv
    -\frac{1}{n} \log\!\int\!\! dQ^\text{\tiny L}dQ^\text{\tiny R}e^{-N (f[Q^\text{\tiny L}]+ f[Q^\text{\tiny R}]) -z \text{Tr}(Q^\text{\tiny L}-Q^\text{\tiny R})^2 }\\
    \text{with\ \ Tr}(Q^\text{\tiny L}-Q^\text{\tiny R})^2=\sum_{a,b=1}^{n}(Q_{a,b}^\text{\tiny L}-Q_{a,b}^\text{\tiny R})^2 \,,
    \label{eq:int1}
\end{multline}
where $Q^L$ and $Q^R$ are $n\times n$ matrices. As in the ferromagnet, such an integral appears when we couple two mean-field models, and ${\cal D}(z,N)\equiv\partial_z{\cal F}(z,N)$ has the physical meaning of a correlation function
\begin{align}
    {\cal D}(z,N)=\mathbb{E}\big(\langle (q^L-q^R)^2\rangle\big) \,,
\end{align}
where $q^L$ and $q^R$ are the overlaps between configurations in the two sub-systems, $\langle\cdot\rangle$ is the Boltzmann average, and $\mathbb{E}(\cdot)$ is the average over the disordered couplings.
It is worth noticing that when ${\cal D}(z,N)$ vanishes, the two subsystems synchronize, i.e., $Q^L=Q^R$, and the whole system of $2N$ spins acquires the ultrametric properties of the common overlap matrix \cite{franz1992ultrametricity, panchenko2015free}. A similar synchronization phenomenon is a key property in off-equilibrium aging dynamics \cite{castillo:02, parisi2003local, kurchan2023time, crisanti2023dynamical}.

In the region where $z=O(1)$ and $N\to \infty$, we find the equivalent of Eq.~\ref{eq:O3}: the mode locking integral, which apart from $z$ -independent prefactor, becomes
\begin{align}
    {\cal Z}(z) \equiv &\sum_\pi \exp \big(-z \; \text{Tr}(Q^*-Q^*(\pi))^2\big)
    \label{eq:int2}
\end{align}
where $Q^*$ is one of the {\sl maxima} of $f[Q]$, $\pi$ is an element of the permutation group $S(n)$, and $Q^*(\pi)$ is the result of the action of permutation $\pi$ on the matrix $Q^*$, that is $Q^*(\pi)_{a,b}=\sum_{c,d}\pi_{a,c}^{-1}Q_{c,d}\pi_{d,b}=Q_{\pi(a),\pi(b)}$.

Two serious problems are present:
\begin{itemize}
    \item In the limit $n\to 0$, the group of symmetry is the permutation group of zero elements. We do not have a convenient way to parametrize it. A direct analytic computation with replicas exceeds our current capabilities. We will write later the function ${\cal F}(z) \equiv \lim_{N\to\infty}{\cal F}(z,N)$ as an integral over an infinite-dimensional space (see Eq.~\ref{eq:quaranta}),  which we effectively approximate as a finite-dimensional integral. 
    \item Even if we compute ${\cal F}(z)$ things are not simple as in the ferromagnetic case. The function $f[Q]$ has zero modes also in directions that do not correspond to the permutation of the matrix. The large $z$ behavior of ${\cal D}(z)$ does not match the small $t$ behavior obtained in perturbation theory for $z=tN.$ Further work is needed.
\end{itemize}
In the Materials and Methods section, we describe how to estimate ${\cal D}(z)$ in the region of large $z$. Indeed, the function ${\cal D}(z)$ may be evaluated using the probabilistic version of the replica approach. We find an explicit expression as the average over two coupled random spin glass trees that we can estimate by randomly generating trees using the Bolthausen-Sznitman (BS) coalescent construction of Ref.~\cite{bolthausen:98}. Although a theoretical estimate of the large-$z$ behavior is attainable using probabilistic tools, we did not attempt it here.
At the end, we find (neglecting possible log terms or small powers) that for large $z$ 
\begin{equation}
{\cal F}(z) \propto z^{1/2} \implies {\cal D}(z) \equiv \frac{d{\cal F}}{dz}\propto z^{-1/2} 
\end{equation}
Let us look at what happens for values of $z$ that grow with $N$ using the notation
$D(t,N) \equiv {\cal D}(z=tN,N)$.
In the limit where $t\equiv z/N$ is finite and small, we can use the perturbative theory around mean field theory and get
\begin{equation}
\lim_{N\to\infty}  N D(t,N) \equiv B(t) \underset{t\to 0}{\approx}  t^{-3/2}
\end{equation}
where, for simplicity, we have neglected a logarithm in the behavior for $t$ near zero. 
The crucial difference from the ferromagnet is that, in this case, the large-$z$ and small-$t$ behaviors do not match as $N\to\infty$. The perturbative formula cannot hold in the region of very small $t$ where $t=O(1/N)$ because it would give a result proportional to $N^{1/2}$ for the quantity $D$ that must be less than 2.

\begin{figure}
    \centering
    \includegraphics[width=\linewidth]{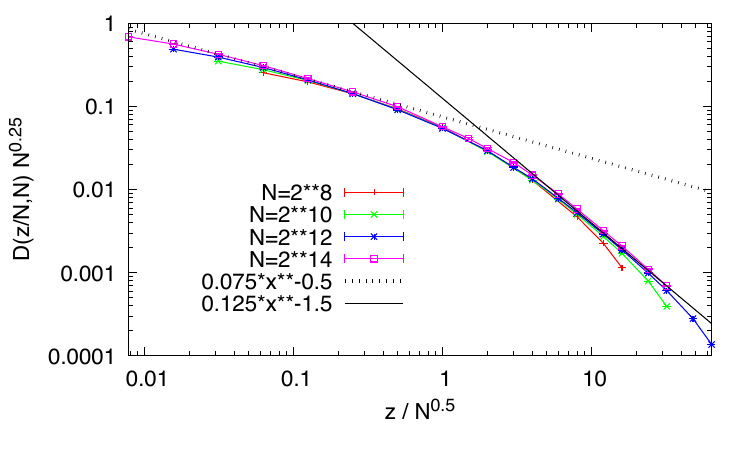}
    \caption{Scaling of the function $D(t=z/N,N)$ according to the theory that predicts the decays $D(t,N) \propto t^{-3/2}$ at fixed $N$ and ${\cal D}(z) \propto z^{-1/2}$. The scaled data in the figure follow the function $\widehat{C}(x)/x^{1/2}$ with $x=z/N^{1/2}$.}
    \label{fig:scaled}
\end{figure}

If we put everything together, in the interesting region where $t\ll 1$, we obtain the main result
\begin{equation}
  D(t,N) \approx {\cal D}(tN) \widehat{C}(tN^{1/2}) = {\cal D}(z) \widehat{C}(z/N^{1/2})\,,
  \label{eq:main}
\end{equation}
where $\widehat{C}$ is a cross-over function such that $\widehat{C}(0)=1$ and $\widehat{C}(x)\propto 1/x$ for large $x$. In this way, we get $D(t,N) \propto 1/(N t^{3/2})$ for $t\to 0$ and ${\cal D}(z)\propto 1/z^{1/2}$ for $z\to\infty$.
The previous formula implies that choosing an $N$-dependent scale 
$t=u/N^\kappa$ we have $D(uN^{-\kappa},N) \propto N^{-\alpha(\kappa)}$ with
\begin{equation}
    \alpha(\kappa)=
     \left\{
     \begin{array}{ll}
        1-\frac32\kappa   & \kappa\le 1/2 \\
         \frac{1-\kappa}{2} & \kappa\ge1/2
     \end{array}
     \right.
    \label{eq:final}
\end{equation}
Both expressions coincide for $\kappa=1/2 $, giving $\alpha=1/4$.

The validity of the crossover scaling functions reported above is supported by the data shown in Fig.~\ref{fig:scaled}, which have been obtained simulating two coupled spin glass models defined on random graphs with different sizes $N$ and coupling strengths $z$. The simulation details are reported in the Materials and Methods section.

\section{Finite size corrections in mean-field spin glasses}

In the $O(3)$ (or more generally $O(n)$) ferromagnetic model, up to logarithmic terms, the finite-volume corrections are simply proportional to $1/N$. The corrections arise from fluctuations in the order parameter around the saddle-point value. The computations are straightforward; they are reported in the SI for the reader's convenience.

In spin glasses, the situation is more interesting.
In the Sherrington-Kirkpatrick (SK) model, finite-size corrections have been computed numerically in a wide range of $N$, the number of spins, up to $N=O(10^4)$. The numerical data for the energy density in the low-temperature region have been fitted to
\begin{equation}
    E(N)=E_\infty +A\, N^{-\omega}\;,
    \label{Crit}
\end{equation}
with a non-trivial exponent $\omega<1$, because the spin glass phase is marginally stable. The Hessian at the saddle point, besides being non-invertible at the critical point, is also non-invertible in the whole low-temperature phase: there is an infinite number of zero modes that accumulate toward zero \cite{de1984spin}. This implies finite size corrections larger than $O(1/N)$ and an exponent $\omega$ that takes a value smaller than $1$. Its determination is the concern of the present section. Previous numerical simulations \cite{boettcher2005extremal, palassini2008ground, boettcher2010simulations, billoire2006numerical} as well as some theoretical attempts \cite{aspelmeier2008finite} estimated $\omega=2/3$; however, this value is not grounded in any solid theory, and an oversimplified computation \cite{parisi2019study} gives $\omega=4/5$.

Here, we relate the computation of $\omega$ to our computation of the mode locking integral.
We show that the value $\omega=2/3$ is incorrect and argue that the correct one is $\omega=3/4$, with possible logarithmic corrections that we systematically neglect.
The resulting asymptotic behavior of the free-energy density is 
\begin{equation}
    f(N)=f_\infty +\frac{A}{N^{3/4}}+\frac{B}{N} \,.
    \label{eq:due}
\end{equation}

As previously exploited by Billoire \cite{billoire2006numerical}, our starting point is the Guerra-Toninelli formula \cite{guerra:02}, which relates finite-size corrections to a generalization of the mode-locking integral, by constructing a model that interpolates between two systems with $N$ sites and one with $2N$ sites.
We write its Hamiltonian as 
\begin{equation*}
    H_t[\sigma_1^{2N}]=\sqrt{1-t}\Big( H_L[\sigma_1^N]+ H_R[\sigma_{N+1}^{2N}]\Big)+\sqrt{t}\,H[\sigma_1^{2N}]\;,
\end{equation*}
where $\sigma_i^k\equiv\{\sigma_i,\ldots,\sigma_k\}$ and
\begin{eqnarray}
H_L[\sigma]&=&N^{-1/2}\sum_{1\le i<j\le N} K^L_{ij}\sigma_i\sigma_j \nonumber \\
H_R[\sigma]&=&N^{-1/2}\sum_{N+1 \le i<j\le 2N} K^R_{ij}\sigma_{i}\sigma_{j} \nonumber \\
H[\sigma]&=&(2N)^{-1/2}\sum_{1 \le i<j\le 2N} J_{ij}\sigma_i\sigma_j\,.
\label{eq:hhh}
\end{eqnarray}
All the couplings $K^L$, $K^R$, and $J$ are independent Gaussian random variables.
If $t=0$, we have a pair of independent SK models with $N$ spins each; while, if $t=1$, we have a single SK model with $2N$ spins.
The partition function is $Z_{2N}(t)=\sum_{\sigma}e^{-\beta H_t[\sigma]}$,
giving a free-energy density $f_{2N}(t)= (2N)^{-1} \mathbb{E}(\log Z_{2N}(t))$, such that $f_{2N}(1)=f_{2N}$ and $f_{2N}(0)=f_{N}$.
Guerra and Toninelli found the  following remarkable identity, instrumental in proving the existence of the thermodynamic limit
\begin{eqnarray}
    && \frac{df_{2N}(t)}{dt}= -\frac{\beta}{4} D(t,N)\;, \nonumber\\
    && 0 \le D(t,N) = \mathbb{E}{ \left\langle (q^{L}-q^{R})^2\right\rangle_t} \le 2\;, \label{eq:GT}\\
    && q^{L}=\frac{1}{N} \sum_{i=1}^{N} \sigma_i\sigma'_i\;, \quad
    q^{R}=\frac{1}{N} \sum_{i=N+1}^{2N} \sigma_i\sigma'_i\;. \nonumber
\end{eqnarray}
where the average $\langle\cdot\rangle_t$ is over two copies $\sigma$ and
$\sigma'$ with Hamiltonian $H_t$, and the expectation is the average over the random couplings.
We get therefore
\begin{equation}
\Delta f = f_{2N}-f_{N} = -\frac{\beta}{4}\int_0^1 dt \; D(t,N) \leq 0\;.
\label{Delta}
\end{equation}
 Introducing replicas to average the free energy over the disorder, we obtain precisely the mode-locking free energy, Eq.~\ref{eq:int1}, with $z=\beta^2 Nt/2$. 

Now, if we can control the behavior of $D(t,N)$ for large $N$ for all values of $t$, we have access to the finite volume corrections.
The existence of the free energy in the thermodynamic limit implies that $D(t,N)$ tends to zero for $N\to \infty$ and fixed $t>0$,
while for $t=0$, due to independence, one has $D(0,N)\to 2\mathbb{E}( \langle q^2\rangle)$.

This is a manifestation of the synchronization (or overlap locking) mechanism, first identified as a crucial feature of the Replica Symmetry Breaking (RSB) phase of spin glasses in Ref.~\cite{franz1992ultrametricity} and later rigorously proven by Panchenko \cite{panchenko2015free}. As soon as the two subsystems are coupled ($t>0$), in the $N\to\infty$ limit, the overlaps in the two subsystems are locked, i.e., for any pair of states $\alpha$ and $\beta$ of the compound system $q_{\alpha\beta}^{L}=q_{\alpha\beta}^{R}$ and $P(q^{L}, q^{R})=\delta(q^{L}-q^{R})P(q^{L})$.
The finite-volume free energy corrections are related to differences in overlaps between the coupled subsystems.  
If we now insert Eq.~\ref{eq:main} in the Guerra-Toninelli expression, we recover Eq.~\ref{eq:due} with an exponent $3/4$.

\begin{figure}
    \centering
    \includegraphics[width=0.9\linewidth]{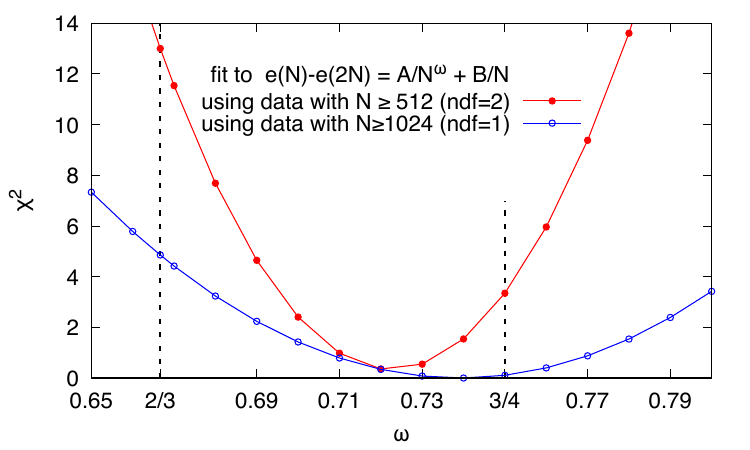}
    \caption{The normalized $\chi^2$ of the fit in Eq.~\ref{eq:fit} to the energy corrections in mean-field spin glasses as a function of the exponent $\omega$. Using the data with $N\ge 512$ and $N\ge 1024$ the analytically predicted value $\omega=3/4$ is statistically acceptable and preferred to previous estimates.}
    \label{fig:chi2}
\end{figure}

To test this prediction, we have measured the energy of mean-field spin-glass models defined on random regular graphs with $N$ vertices and fixed degree 4. The $\omega$ exponent should be the same as in the SK model, as far as the structure of the tree of states should be the same \cite{panchenko2015hierarchical,panchenko2016structure}, but a sparse model can be simulated much more efficiently. We can thermalize systems of size up to $N=8192$ well within the low-temperature spin-glass phase (down to $T=T_c/2$). We simulate a huge number of samples (130k for $N=512$, 400k for $N=1024$, 700k for $N=2048$, 200k for $N=4096$, 51k for $N=8192$) to obtain a relative uncertainty on the mean intensive energy $e(N)$ of a few parts in a million, much smaller than in previous computations. We interpolate data with the law
\begin{equation}
    e(N) - e(2N) = A / N^\omega + B / N\;,
    \label{eq:fit}
\end{equation}
for several values of $\omega$ and report in Fig.~\ref{fig:chi2} the $\chi^2$ value of the fit as a function of $\omega$.  The presence of the $1/N$ corrections strongly increases the error in the fits, however its presence is predicted by the theory. We discard data with $N<N_\text{min}$ and use two values $N_\text{min}=512,1024$. We notice that the value $\omega=2/3$ is never statistically acceptable, while the value $\omega=3/4$ is not only perfectly acceptable, but becomes the most probable value when the analysis is restricted to the largest sizes.

\section{Correlations in the Hierarchical Model}

A long-standing problem in finite-dimensional spin glasses is the determination of the distance dependence of the overlap connected correlation function $\mathbb{E}(\langle q(x)q(y)|q\rangle_c)= \mathbb{E}(\langle q(x)q(y)|q\rangle)-q^2$ for fixed value of the total overlap $q$ \cite{de1984spin,contucci:09,janus:10}. This has a large distance power law behavior $\mathbb{E}(\langle q(x)q(y)|q\rangle_c) \simeq |x-y|^{-\alpha(q)}$ with an exponent $\alpha(q)$ which depends of $q$.
We analyze here Dyson hierarchical spin glasses and show that the exponent $\alpha(q)$ exhibits nontrivial behavior due to non-perturbative effects.

The overlap-overlap coupling between the systems studied in the previous sections lies at the heart of the definition of Dyson hierarchical spin-glass models.  The Hamiltonians of these models are constructed recursively, coupling systems of size $N$ to form systems of size $2N$.
Starting with free spins for $N=1$, one defines 
\begin{multline}
    H_{2N}[\sigma] = H_{L,N}[\sigma^L] + H_{R,N}[\sigma^R] + \Delta H_N[\sigma^L,\sigma^R]\\
    \text{with}\quad\Delta H_N[\sigma^L,\sigma^R] = N^{-\rho/2}\sum_{i\in L,\; j\in R} J_{ij} \sigma_i^{L} \sigma_j^{R}
\end{multline} 
where the $J_{ij}$ are standard Gaussian variables. The exponent $\rho\in [1,2]$ defines the properties of the model: for $1<\rho<\frac{4}{3}$ the model is in the mean-field universality class, while for $\frac{4}{3}<\rho< 2$ the system has anomalous critical behavior corresponding to an effective spatial dimension $D_\text{eff}=2/(\rho-1)$, in the same sense as $D=1$ long-range model may approximate a short-range model in $D_\text{eff}$ spatial dimensions \cite{kotliar:85, banos:12,angelini2014relations, jensen:21}. The extrema of this interval correspond to effective dimensions $D_\text{eff}=6$ and $D_\text{eff}=2$, respectively.

Briefly, the interaction Hamiltonian is a random Gaussian quantity with covariance given by
\begin{equation*}
    \mathbb{E}(\Delta H_N[\sigma_\text{\tiny L},\sigma_\text{\tiny R}]
    \Delta H_N[\sigma'_\text{\tiny L},\sigma'_\text{\tiny R}])=
    N^{2-\rho} q[\sigma_\text{\tiny L},\sigma'_\text{\tiny L}] q[\sigma_\text{\tiny R},\sigma'_\text{\tiny R}]
\end{equation*}
The interaction term between two subsystems is a sub-extensive random quantity of order $N^{1-\rho/2}$ 
giving an energy contribution of order $N^{2-\rho}$.
Let us compute the correlations of the overlap at distance $N$, i.e., at the highest level of the hierarchy
\begin{equation}
   \langle (q_L-q_R)^2 \rangle_N=D( N^{1-\rho},N)\,,
\end{equation}
where $q_L=q(\sigma_L,\sigma'_L), q_R=q(\sigma_R,\sigma'_R)$ are the overlaps in the two subsystems.

Let us hypothesize that the structure of the fluctuations of $q_L-q_R$ is similar to that of the SK model. In this case, using Eq.~\ref{eq:final} we find 
\begin{equation}
     D( N^{1-\rho},N)\propto \left\{
    \begin{array}{ll}
    {N^{-(5-3 \rho)/2}} & \rho<3/2\\ 
    {N^{-(1-\rho/2)}} & \rho>3/2
    \end{array}
    \right.
\end{equation}
For $\rho<3/2$, we have the result of naive perturbation theory, while for $\rho>3/2$, the result is corrected by the presence of the crossover and is of a non-perturbative nature. 
Interestingly, we can generalize the analysis to study the correlation function at fixed average overlap $q$. Introducing $q=(q_L+q_R)/2$ and 
$s=q_L-q_R$ we can restrict the function $D$ to fixed $q$ and define 
$D(t,N|q)=\mathbb{E}(\langle s^2|q\rangle_{t,N}).$ For $t= N^{1-\rho}$ this correlation should decay as a power of the distance $N$, $D(N^{1-\rho},N|q)=A(q)/N^{\alpha(q)}$ with an exponent $\alpha(q)$ that depends on $q$. $D(t,N|q)$ is a proxy for $\mathbb{E}(\langle q(x)q(y)|q\rangle_c)$ on distances $|x-y| \sim N$, as explicitly shown in the SI.

In the small $t$ region for $N\to\infty$, $D(t,N|q)$ is given by the perturbation theory \cite{de1985gaussian, temesvari1988long}, which gives a different result for $q$ equal or different from zero and $q_\text{\tiny EA}$:
\begin{equation}
    D(t,N|q)\propto\left\{
    \begin{array}{ll}
    (Nt^2)^{-1}     & q=0 \\
    (Nt^{3/2})^{-1} & 0<q<q_\text{\tiny EA}\\
    (Nt)^{-1}       & q=q_\text{\tiny EA}
    \end{array}
    \right.   
    \label{eq:26}
\end{equation}
We conjecture now that, conversely, in the region of finite $z=Nt=O(1)$ the function ${\cal D}(z|q)$ has the same behaviour as the unconditional function ${\cal D}(z)$, i.e., ${\cal D}(z|q) \simeq {\cal D}(z) \simeq z^{-1/2}$ for large $z$. As before, the matching should be provided by a crossover function.
The explicit verification of this hypothesis is not straightforward, and in its favor, we have only moderate numerical evidence. We present it because it yields important consequences that are in good agreement with numerical simulations.
If we fix $0<q<q_\text{\tiny EA}$ strictly, the crossover is the same as for the unconditional correlation, and nothing special happens
\begin{equation}
    \alpha(q)=\max\left(\frac{5-3\rho}{2},1-\frac{\rho}{2}\right)\,.
\end{equation}
For $q=0$, instead, the crossover is from $z^{-1/2}$ to $1/(Nt^2)$ that gives 
\begin{equation}
    D( N^{1-\rho},N|q=0)\propto \left\{
    \begin{array}{ll}
    {N^{-(3-2 \rho)}} & \rho<4/3\\ 
    {N^{-(1-\rho/2)}} & \rho>4/3
    \end{array}
    \right.
\end{equation}
and consequently 
\begin{equation}
    \alpha(0)=\max\left(3-2 \rho, 1-\frac{\rho}{2}\right).
\end{equation}

\begin{figure}
    \centering
    \includegraphics[width=0.8\linewidth]{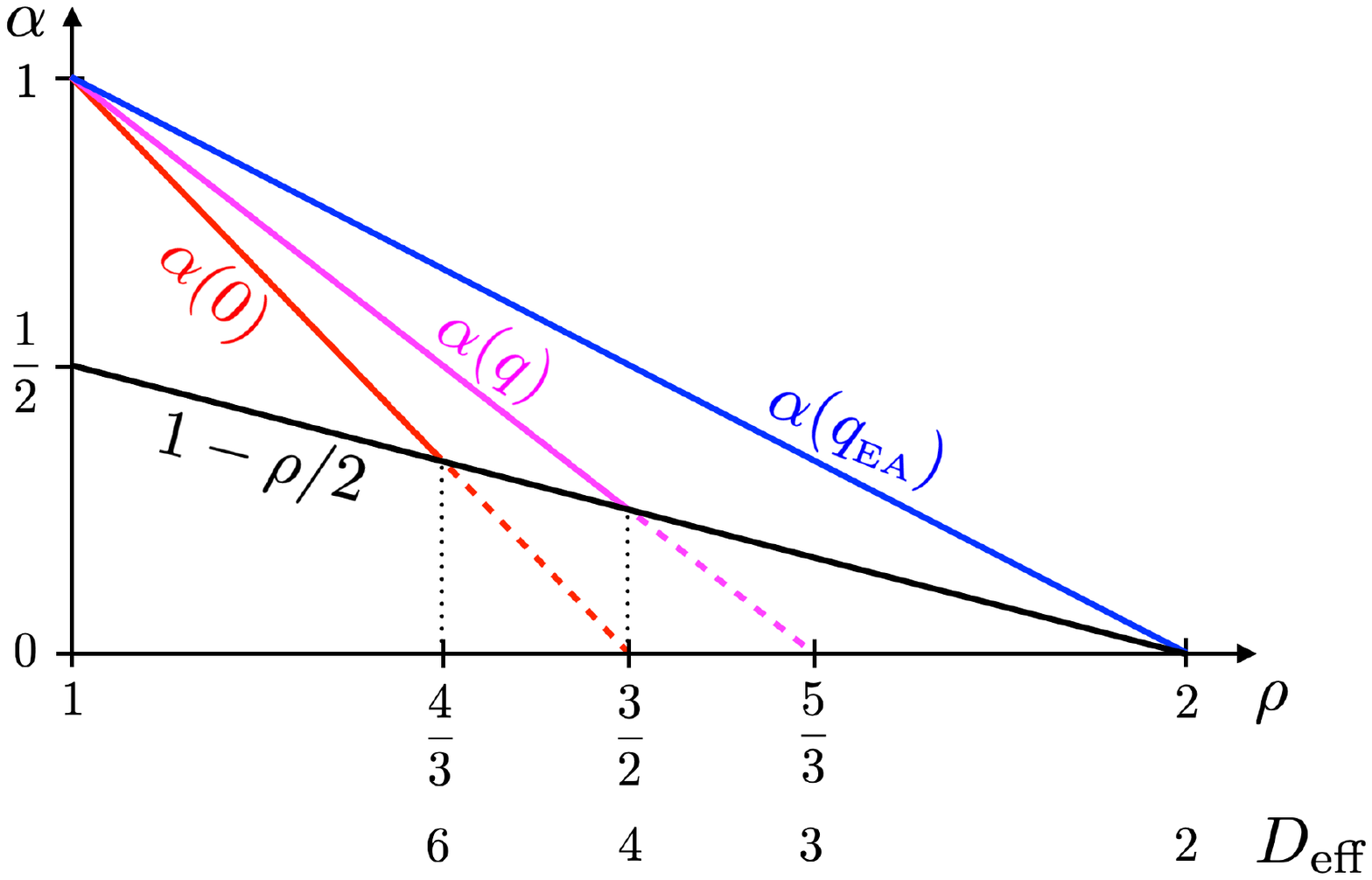}
    \caption{Our prediction for the exponent $\alpha(q,\rho)$ in the one dimensional long range model as function of $\rho$}
    \label{fig:esponenti}
\end{figure}

Remarkably, this is the behavior found numerically for the long-range one-dimensional model in the very interesting paper \cite{jensen:21}, where the exponent $\alpha(0)$ has been computed in asymptotic off-equilibrium dynamics, which corresponds to the global $q=0$ sector. The hierarchical model is a proxy for the long-range one-dimensional model, so these predictions could also be applied to it.
Notice that when $\rho\geq 3/2$ the two exponents $\alpha (q)$ and $\alpha(0)$ collapse: this is exactly what happens in dimensions $D=3$ for short range models where $\alpha(0)=\alpha(q)$ for $q<q_\text{\tiny EA}$ \cite{janus:10}. For $q=q_\text{\tiny EA}$, the exponents likely remain equal to their perturbative values $\alpha(q_\text{\tiny EA})=2-\rho$, which coincide with the \emph{critical} decay exponents for any $\rho<2$ \cite{kotliar:85}. The results are depicted in Fig.~\ref{fig:esponenti}.
These results are non-perturbative in nature because they come from the region where $t$ goes to zero as a function of $N$, while perturbative corrections, unfortunately unknown at the present time, should change the behavior at small $t$ at $N=\infty$. When perturbative renormalization group computations will be available, if we denote by $\alpha^*$ the results of these computations, we conjecture that the non-perturbative lower bound $\alpha(\rho) \ge 1-\rho/2$ remains valid, and the following hold
\begin{align}
  \alpha(q)&=\max\left(\alpha^*(q,\rho), 1-\frac{\rho}{2}\right)\,, \\
  \alpha(0)&=\max\left(\alpha^*(0,\rho), 1-\frac{\rho}{2}\right)\,.
\end{align}

In conclusion, we have obtained, albeit with some conjecture, the formulae for the exponent of the correlation functions by performing non-perturbative scaling computations. This may be surprising, but it is well known that, in the long-range model (and also in the hierarchical model), the exponents that control the large-distance decay of the order-parameter correlations are not renormalized perturbatively. The fact that the exponents differ from the mean-field values has always been puzzling. Here, we find that the resolution of the conundrum is simple: the difference between the exponent and the naive value is a non-perturbative effect that can be mastered analytically.

\section*{Acknowledgements}
We are grateful to Alain Billoire for useful discussions.
The research has received financial support from the “National Center for HPC, Big Data and Quantum Computing”, Project CN\_00000013, CUP B83C22002940006, NRRP Mission 4 Component 2 Investment 1.4,  Funded by the European Union - NextGenerationEU.

\matmethods{
\subsection*{The exponent of the finite volume corrections}
Let us prove the formula in Eq.~\ref{eq:main}.
We go through the following three steps.

(1) At fixed $t$ and large $N$,  the coupling of the two systems is extensive and
\begin{equation}
    D(t,N) \approx B(t)/N +B_1(t)/N^2 +.... \label{eq:Pert}
\end{equation}
where $B(t)$ is a function that has been computed in perturbation theory in \cite{de1984spin}; it diverges for $t\to 0$ as $t^{-3/2}$ at small $t$ (more precisely as $t^{-3/2}(a+b\log(t))$, but we are neglecting logarithms). The function $B_1(t)$ is, in principle, computable, but the computations are too complex to be done.
This result implies the following scaling behavior $D(t,N)=A/(Nt^{3/2})$, suggesting that $D$ would become of $O(1)$ at the scale $t\sim N^{-2/3}$.
That would agree with the numerical results of Billoire \cite{billoire2006numerical}, who found that in a given region of parameters $D(t,N)=S(tN^{2/3})$ with $S(x) \approx A/x^{3/2}$ for large $x$.

(2) This scaling cannot hold exactly as it suggests that locking takes place on a scale $t\sim N^{-2/3}$. By general arguments, a finite energy disturbance to the Hamiltonian---corresponding to finite $z$---reshuffles all the weights of the equilibrium states. The function ${\cal D}(z)$ is a decreasing function of $z$, and should go to zero for $z\to\infty$. Locking takes place on the scale $t\sim N^{-1}$ rather than $t\sim N^{-2/3}$. 

In the following, we will compute ${\cal D}(z)$ in a power expansion for small $z$ and by a semi-analytical statistical method for large $z$. 

(3) For large $z$ we find semi-analytically and numerically that ${\cal D}(z)$ has a power law behavior 
\begin{equation}
  {\cal D}(z)\approx a z^{-\gamma}
\end{equation}
and we estimate $\gamma=1/2$.
Such behavior is incompatible with the Billoire scaling $D(t,N)=S(t N^{2/3})$ in the region where $z=O(1)$, as it would lead to a constant ${\cal D}(z)$.
We can match the two regimes
\begin{itemize}
    \item $D(t,N) \sim (t N)^{-1/2}$ for large $tN$
    \item $t=O(1)$ where for small $t$, $D(t,N)\sim 1/(t^{3/2} N)$, where the Billoire scaling is satisfied
\end{itemize}
only assuming a smooth interpolation in the region $t\sim N^{-1/2}$ where the two are of the same order of magnitude. We write the scaling form as
\begin{eqnarray}
    D(t,N)\approx {\cal D}(Nt)\;C(N^{1/2}t) 
    \label{Fin}
\end{eqnarray}
where the cross-over function $C(y)$ tends to one at small $y$ and behaves as $1/y$ at large $y$. 

In this way, in the region where both arguments are large, we get the behavior $D(t,N)\propto (Nt)^{-1/2}/(N^{1/2}t) =1/(Nt^{3/2})$ observed in the simulations. 
Notice that this scaling depends on the assumption $\gamma=1/2$; different values of $\gamma$ would give a different scaling. There is, however, no way of getting $\alpha=2/3$ if $\gamma>0$. 
Unfortunately, the crossover function $\widehat C(y)$ does not appear to be computable in a straightforward manner. In principle, we could use Eq.~\ref{eq:Pert}: our scaling implies that $B_1(t)\approx b_1/t^{7/2}$, but computation of $B_1(t)$ is extremely complex (an approximate computation taking care of only the longitudinal fluctuations is described in Ref.~\cite{parisi1992finite,parisi2019study}).

\subsection*{The computation of ${\cal D}(z)$}
Let us now turn on the computation of the function ${\cal D}(z)$, using the probabilistic approach to spin glasses.

In the region of large $N$ and fixed $z=Nt$ we can rewrite the partition function of the weakly coupled systems as follows
\begin{multline}
\widehat Z_{2N}(z)\equiv Z_{2N}\big(t=\frac{z}{N}\big)=
\sum_{\{\sigma,\tau\}}e^{-\beta\big(\widetilde{H}_{L,z}[\sigma] + \widetilde{H}_{R,z}[\tau] + \sqrt{z}\,\Delta[\sigma,\tau]\big)}\\
\text{with}\quad \widetilde{H}_{K,z}[\tau]=\big(1-\frac{z}{2N}\big)H_K[\tau]+\sqrt{\frac{z}{2N}}H'_K[\tau] \quad (K=L,R)\\
\text{and}\quad \Delta[\sigma,\tau]=N^{-1/2}
\bigg(H[\sigma,\tau]-\frac{1}{\sqrt{2}}\big(H'_L[\sigma]+H'_R[\tau]\big)\bigg)
\label{eq:inter}
\end{multline}
where $H_K[\tau]$ and $H'_K[\tau]$ are independent samples with the same covariance $\mathbb{E}(H_K[\tau]H_K[\tau'])=\mathbb{E}(H_K'[\tau]H_K'[\tau'])=\frac{N}{2}q[\tau,\tau']^2$, being $q[\tau,\tau']\equiv N^{-1}\sum_{i=1,N}\tau_i \tau'_i$.
Notice that $H_{K,z}$ includes $O(1)$ corrections to $H_K$.
The interactions term $\Delta[\sigma,\tau]$ is an $O(1)$ correction to the independence of the two systems; more precisely, it is a random Gaussian variable with covariance $\mathbb{E}({\Delta[\sigma,\tau]\Delta[\sigma',\tau']})=q[\sigma,\sigma']q[\tau,\tau']$.
Therefore, the function $\widehat Z_{2N}(z)$ can be computed from the known probability distribution of the overlaps.

Let us see how it works. We recap the theory of spin glasses in simplified terms. In the RSB phase, in each instance of the system, there is an infinite number of equilibrium states labeled by $\alpha$. These states have a statistical weight $w_\alpha$, and mutual overlaps $q_{\alpha,\gamma}$.  
The weights satisfy the condition $\sum_\alpha w_\alpha =1$ and  $q_{\alpha,\gamma}=\frac{1}{N} \sum_{i=1}^{N}  \langle \sigma_i\rangle_\alpha \langle\sigma_i\rangle_\gamma$, where $\langle \cdot \rangle_\alpha$ denotes the statistical average in the state labeled by $\alpha$. The probability distributions of these quantities are well known and will be described later.

Consider now the partition function in Eq.~\ref{eq:inter}. At $z=0$ we have two independent subsystems.
The set of compound states is the Cartesian product of the states of each subsystem, which we label by $(\alpha\gamma)$.
The weights will just be given by the products $W_{\alpha\gamma}(0)=w_\alpha^L \; w^R_\gamma$.
With transparent notation, 
\begin{equation}
    \mathbb{E}\left({\langle q^2\rangle}\right)=
    \mathbb{E}\left(
    \sum_{\alpha,\alpha',\gamma,\gamma'}
    w_\alpha^L w_{\alpha'}^L w^R_\gamma w^R_{\gamma'}
    \left( \frac{q_{\alpha,\alpha'}^L+q^R_{\gamma,\gamma'}}{2}\right)^2\right)
\end{equation}
and
\begin{equation}
 {\cal D}(0)=
  \mathbb{E}\left(
  \sum_{\alpha,\alpha',\gamma,\gamma'}
  w_\alpha^L w_{\alpha'}^L w^R_\gamma w^R_{\gamma'}
  \left( q^L_{\alpha ,\alpha '}-q^R_{\gamma ,\gamma'}\right)^2\right)\,.
\label{eq:d0}
\end{equation}

The average interaction energies between the two subsystems, $\Delta_{\alpha,\gamma}$, are random Gaussian variables with covariance
\begin{equation}
\mathbb{E}\left(\Delta_{\alpha,\gamma}\Delta_{\alpha',\gamma'}\right)=q^L_{\alpha,\alpha'}q^R_{\gamma,\gamma'}\,.
\end{equation}
In the presence of such a finite perturbation, the weights of the states 
get modified to
\begin{eqnarray}
    W_{\alpha\gamma}(z) =
    \frac{w_\alpha^L(z) w^R_\gamma(z) e^{-\sqrt{z} \beta \Delta_{\alpha\gamma}}}{\sum_{\alpha\gamma}w_\alpha^L(z) w^R_\gamma(z) e^{-\sqrt{z} \beta \Delta_{\alpha\gamma}}}\,.
\end{eqnarray}
A remarkable result of replica theory is that the \emph{distribution} of the weights $w_\alpha^K(z)$ is independent of $z$. When computing disorder averages at fixed $z$, we can simply ignore the dependence of $w_\alpha^K(z)$ on $z$.
This is quite general. However, in the absence of the magnetic field, we have an extra symmetry at $z=0$ due to the possibility of reversing all the spins of only one of the two subsystems.
It is thus convenient to consider only positive overlaps, considering together pairs of states in each subsystem that are related by spin inversion. 
With that proviso, the formula becomes 
\begin{eqnarray}
    W_{\alpha\gamma}(z) 
    = \frac{w_\alpha^L w^R_\gamma \cosh\left(\sqrt{z} \beta \Delta_{\alpha\gamma}\right)}{\sum_{\alpha\gamma}w_\alpha^L w^R_\gamma \cosh\left(\sqrt{z} \beta \Delta_{\alpha\gamma}\right)}
\end{eqnarray}
Inserting this expression in Eq.~\ref{eq:d0} we find
\begin{eqnarray}
    {\cal D}(z) = \mathbb{E}\left(\sum_{\alpha\gamma\alpha'\gamma'}
    W_{\alpha\gamma}(z) W_{\alpha'\gamma'}(z)
    \left(q_{\alpha,\alpha'}^L - q^R_{\gamma,\gamma'}\right)^2
    \right)
    \label{eq:quaranta}
\end{eqnarray}
Although the previous formulae are explicit, we cannot obtain the large-$z$ behavior analytically, and we must \textit{resort to the indignity of numerical evaluations}, extracting the $w$'s and $q$'s from their known probability distributions, as shown in the SI. 
We evaluate the function ${\cal D}(z)$ semi-analytically by generating random trees of spin-glass states. 
We follow here a strategy similar to the one recently used in Ref.~\cite{aguilar2024small} to study chaos in a small magnetic field. 
We use the Bolthausen-Sznitman construction \cite{bolthausen:98} to generate numerically a large number of random trees with $M$ leaves, compute the averages, and extrapolate to the limit $M\to\infty$. Interestingly, to make such a calculation, we only need as input the function $P(q)$, its primitive $x(q)$, or its inverse $q(x)$. The range of the values of the variable $x$ is $0\le x<x_M$, where $q(0)=0$ and $q(x_M)=q_\text{\tiny EA}$. In the numerics, we will consider the very simple choice $q_\text{\tiny EA}=0.4$, $\beta=1.5$, $x(q)=q/3$ (i.e., $q(x)=3x$, that is the form of the SK near $T_c$).

We first describe how the computation is performed in principle, and later, the techniques we used to reduce computational time. In Ref.~\cite{aguilar2024small}, it was shown how to generate random trees with $M$ leaves.
The algorithm produces a list of the weight of the $M$ leaves and the corresponding $M(M-1)/2$  overlaps between pairs of leaves.
It also provides the tree branching points that correspond to different clusters of states.
These branching points are labeled by a real variable $x$ that is related to the value of the overlap between the states that belong to that cluster by the function $q(x)$. The variable $x$ ranges from $x=0$ at the root and $x=x_M$ at the leaves. In the following, we refer to nodes as both branching points and leaves (terminal nodes).

Our system is composed of two separate subsystems, and therefore its states can be represented as the direct product of two random trees. Once a pair of random trees with the corresponding weights is generated,  we construct their direct product. We can associate the Gaussian energies $\Delta_{\alpha,\gamma}$ with the leaves according to the procedure described in the SI.
This allows us to compute $\langle q_L q_R\rangle_\text{single}$ for a single system. Averaging over the trees allows the computation of ${\cal D}(z)$. A few comments are in order:
\begin{itemize}
    \item The computation should be done in the limit where the number $M$ of leaves of each tree goes to infinity. In our study, we consider the cases $M=10^2,10^3,10^4,10^5,10^6,10^7$. In our code, the computation time and memory usage scale approximately as $M^{1.5}$, so we extrapolate to $M=\infty$ from these values.
    \item We can compute for an arbitrary form of the function $q(x)$. Here we use $q(x)=3 x$, $x_M=2/15$, $T=0.7$, which mimics the values of the function $q(x)$ for the SK model at that value of the temperature.
\end{itemize}

\begin{figure}[t]
\centering
\includegraphics[width=0.9\linewidth]{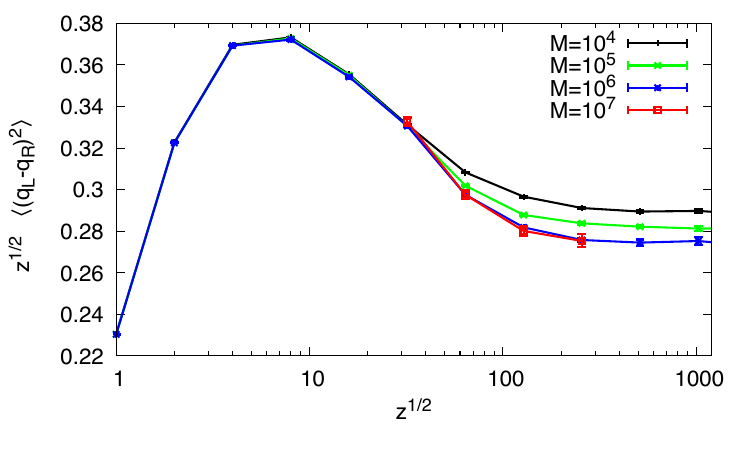}
 \caption{The semi-analytic predictions for $z^{1/2}{\cal D}(z)$ plotted versus $z^{1/2}$ for $M=10^4,10^5,10^6,10^7$. A constant behaviour at large $z$ of this quantity implies that ${\cal D}(z)\propto z^{-1/2}$.}
 \label{fig:delta}
\end{figure}

We show these semi-analytic results for the function ${\cal D}(z)$ in Fig.~\ref{fig:delta}. The dependence on $M$ becomes stronger when $z$ increases, and it is negligible for not too large $z$. Indeed, the reweighting factors increase exponentially with $z^{1/2}$: in the large-$z$ region, states with asymptotically small values of the weights become dominant. However, with this proviso, it seems that for large $M$ the $M$-dependence becomes weak and for large values of $z$ the quantity $z^{1/2}{\cal D}(z) $ is nearly constant.

Can we do the same computation directly with replicas? As we show in the SI, the function ${\cal D}(z)$ can be computed with the replica method as an expansion in powers of $z$. To the first order $z$, we find 
\begin{align}
 {\cal D}(z)=2\langle q^2\rangle-2 z\beta^2 \langle q^2\rangle^2+O(z^2)
\end{align}
where $\langle q^k\rangle=\int dq\; P(q) \; q^k$ and $P(q)$ is the average overlap distribution. In principle, one can compute each order of the expansion in $z$; however, we need a high number of them to extract the large $z$ behavior. This may be possible, but we do not address it in this paper. We have checked with high accuracy that this expansion is in very good agreement with the results of the semi-analytic computation.

\subsection*{The computation of $B(t)$}
The function $D(t,N)$ in the regime of $t=O(1)$ can be studied perturbatively with the replica method. 
A simple computation (following the approach of Ref.~\cite{brunetti1992asymmetric}) shows that the $n$ times replicated partition function can be written in terms of the $n\times n$ overlap matrices $Q^L_{ab}$ and $Q^R_{ab}$ of the two subsystems as  
\begin{multline}
Z_{2N}^{(n)}(t) = \int dQ^L dQ^R 
e^{-\frac{\beta^2}{4} N t \text{Tr}(Q^L-Q^R)^2 - N \beta [F(Q^L)+F(Q^R)]}\\
e^{-\beta f[Q]} = e^{-\frac{\beta^2}{4}\text{Tr}Q^2} \sum_{\{\sigma_a\}}
e^{\beta\sum_{a<b}Q_{ab}\sigma_a\sigma_b}\,.
\end{multline}

The leading corrections at $t=O(1)$ 
can be obtained supposing $\delta Q=Q_1-Q_2$ small, expanding to the second order and performing the Gaussian integration in $\delta Q$. 
In this way, we obtain
\begin{equation}
    Z_{2N}(t)=Z_{2N}(1)\exp\left(-\frac{\beta}{2} \text{Tr}(\log(  {\cal H}+\beta t))\right)\,,
\end{equation}
where ${\cal H}$ is the Hessian matrix with elements ${\cal H}_{ab,cd}=T F''[Q]_{ab,cd}$.
It is well known that severe zero modes plague the spin-glass Hessian; however, the term $\beta t$ regularizes it, and 
\begin{eqnarray}
    B(t)=\frac{d}{dt}\; \lim_{n\to 0} \frac{1}{2n} \text{Tr}(\log({\cal H}+\beta t))
\end{eqnarray}
remains finite for $t>0$.

Zero modes imply that $B(t)$ diverges at small $t$. The consequences of zero modes for the spin-glass propagators at the RSB saddle point were studied long ago by De Dominicis and Kondor \cite{de1984spin}. We can adapt their results to compute $B(t)$, noting that $t$ plays, in our case, the role of the momentum-squared $k^2$ in a finite-dimensional theory. 
Defining $q=(q_L+q_R)/2$ and $s=q_L-q_R$, their theory allows us to compute also the conditional averages of $s^2$ for fixed values of $q$ and small $t$
\begin{eqnarray}
  N D(t,N|q) = N \mathbb{E} \langle s^2|q \rangle_t \simeq \frac{C_1}{t(t+a\,q^2)}+\frac{C_2(q)}{t^{3/2}(t^{1/2}+b\,q)}
  \label{DK2}
\end{eqnarray}
The first term is relevant only when $q$ is near zero. For small $t$ it is (in distribution sense) $C_1\delta (q)/t^{3/2}$. The second term at $q$ not too  small is $ C_2(q)/(q t^{3/2})$, where the function $C_2(q)$ has a non-zero limit when $q\to 0$ and  vanishes linearly at $q=q_\text{\tiny EA}$. Exactly at $q=q_\text{\tiny EA}$ we have:
\begin{eqnarray}
  N \langle (q^{L}-q^{R})^2|q_\text{\tiny EA}\rangle_t\simeq \frac{C_3}{t}
  \label{DK3}
\end{eqnarray}
For $q>q_\text{\tiny EA}$, $N \langle (q^{L}-q^{R})^2|q\rangle$ remains finite when $t$ goes to zero.

The total propagator at fixed $t$, $B(t)$,  is obtained integrating over the values of $q$ with the respective probabilities, and for small $t$, is dominated by the small $q$ contribution, leading to 
\begin{equation}
    \text{Tr}(({\cal H}+\beta t)^{-1})\propto t^{-3/2}\log (t)\,, \quad
    \text{Tr}(\log({\cal H}+\beta t)) \propto t^{-1/2}\log (t)\,.
\end{equation}
Neglecting logs, this result implies the $t^{-3/2}$ divergence of the function $B(t)$ at small argument.
In what follows, the logarithm will be neglected, and we concentrate on power-law singularities.

\subsection*{Direct simulations of the mode-locking integral}
We aim to verify the theory derived above, through numerical simulations, particularly the various scalings predicted.
Unfortunately, numerical investigations in the fully connected model are very time-consuming. They have been done in Ref.~\cite{billoire2006numerical}. Here, in order to increase the accuracy, we introduce an equivalent model defined on sparse random graphs that can be simulated efficiently \cite{aguilar2024small}. The scaling should be the same in the two models.

We simulate a spin glass model defined on a \emph{rewired} sparse random graph, that interpolates between two sparse random graphs of size $N/2$ for $t=0$ and a single sparse random graph of size $N$ for $t=1$ \cite{franz2003replicaI, franz2003replicaII}. For algorithmic convenience, we keep the degree of this family of rewired sparse random graphs constant.

The construction of a $z$-rewired $d$-regular random graph ($d$-RRG) of size $N$ goes as follows. We begin constructing two $d$-RRGs, each of size $N/2$. We use normal indices $i,j\in[1,N/2]$ and primed indices $i',j'\in[N/2+1,N]$ to refer to nodes in the first and second graphs, respectively. Let $E_L$ and $E_R$ denote the edge sets of these two graphs. Initially, the set $I$ of edges $(i,i')$ interconnecting the two graphs is empty. Then we apply $z$ times the rewiring procedure, which removes an edge $(i,j)$ from $E_L$ and an edge $(i',j')$ from $E_R$ at random, and adds the rewired edges $(i,j')$ and $(i',j)$ to the set $I$.
The edge set of the $z$-rewired $d$-RRG of size $N$ is $E=E_L\cup E_R \cup I$.

Given that the maximum number of edges that can be rewired is $z_\text{max}=dN/4$, i.e., the size of both $E_1$ and $E_2$, we define $t=z/z_\text{max}$. Simulations were performed with $d=4$, and thus $t=z/N$ in the following.

We have simulated spin glass models with random binary coupling ($J_{ij}=\pm 1$) on $z$-rewired 4-RRG of sizes $N=2^6,2^8,2^{10},2^{12}$. The number of samples ranges from approximately $10^6$ for $N=2^6$ to $4\,10^3$ for $N=2^{12}$. We have used the Parallel Tempering algorithm to thermalize the systems down to $T=T_c/2$, where $T_c=1/\text{atanh}(1/\sqrt{d-1})\simeq 1.51865$ for $d=4$.
The most relevant observable we have measured is the variance of the overlap difference between the two coupled subsystems
\begin{equation}
D(t,N) = \mathbb{E}\left({\left\langle \left(q_{L}- q_{R}\right)^2\right\rangle_t}\right)\;.
\end{equation}

\begin{figure}
    \centering
    \includegraphics[width=0.9\linewidth]{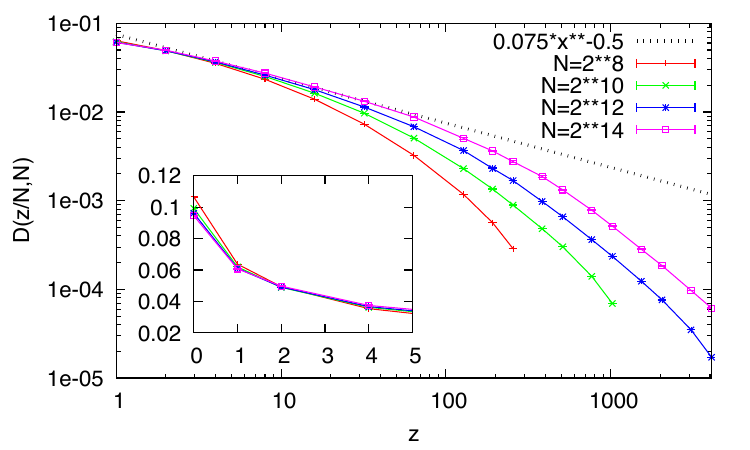}
    \caption{Data for $D(t=z/N,N)$ as a function of $z$ for several system sizes. The dashed line is the predicted power behavior for ${\cal D}(z)=\lim_{N \to \infty} D(z/N,N)$.
    In the insert, we show the same data zoomed into the small-$z$ region, where size dependence is nearly absent, and the function ${\cal D}(z)$ can be measured directly.}
    \label{fig:allz}
\end{figure}

In Fig.~\ref{fig:allz} we show the data $D(t=z/N,N)$ as a function of $z$ for several system sizes, for all the values of $z$, up to $z=N$, where half of the links have been rewired. The dashed line is the predicted behavior in the large $N$ limit, i.e., a power decay $z^{-1/2}$ that is very well compatible with the data. In the insert, we zoom in on the small-$z$ region, where we observe that the large-$N$ limit is reached for not too large values of $N$. 

The main numerical result has already been shown in Fig.~\ref{fig:scaled}, where data have been collapsed following the two scalings predicted by our theory: $D(t,N) \propto t^{-3/2}$ at fixed $N$ and ${\cal D}(z) \propto z^{-1/2}$.
The former scaling is in perfect agreement with that reported in \cite{billoire2006numerical}. The figure also reports the explicit form for the smooth crossover function $\widehat{C}(x)/x^{1/2}$ with $x=z/N^{1/2}$ that is relevant in the region where the above two asymptotic functions do intersect.  In conclusion, our theoretical analysis is perfectly in agreement with the measured behavior of $D(t,N)$.
}

\showmatmethods{}

\bibliography{biblio}

\begin{thebibliography}{10}

\bibitem{parisi:88}
G Parisi, {\em Statistical Field Theory}.
\newblock (Addison-Wesley), (1988).

\bibitem{zinn2021quantum}
J Zinn-Justin, {\em Quantum field theory and critical phenomena}.
\newblock (Oxford university press) Vol.{} 171, (2021).

\bibitem{de1984spin}
C de~Dominicis, I Kondor, On spin glass fluctuations.
\newblock {\em\protect\JournalTitle{Journal de Physique Lettres}} \textbf{45},
  205--210 (1984).

\bibitem{temesvari1988long}
T Temesvari, I Kondor, C De~Dominicis, Long-wavelength fluctuations in the
  ising spin glass.
\newblock {\em\protect\JournalTitle{Journal of Physics A: Mathematical and
  General}} \textbf{21}, L1145 (1988).

\bibitem{franz1994interfaces}
S Franz, G Parisi, MA Virasoro, Interfaces and louver critical dimension in a
  spin glass model.
\newblock {\em\protect\JournalTitle{Journal de Physique I}} \textbf{4},
  1657--1667 (1994).

\bibitem{astuti2019new}
V Astuti, S Franz, G Parisi, New analysis of the free energy cost of interfaces
  in spin glasses.
\newblock {\em\protect\JournalTitle{Journal of Physics A: Mathematical and
  Theoretical}} \textbf{52}, 294001 (2019).

\bibitem{franz2009overlap}
S Franz, T J{\"o}rg, G Parisi, Overlap interfaces in hierarchical spin-glass
  models.
\newblock {\em\protect\JournalTitle{Journal of Statistical Mechanics: Theory
  and Experiment}} \textbf{2009}, P02002 (2009).

\bibitem{wilson:74}
KG Wilson, J Kogut, The renormalization group and the $\epsilon$-expansion.
\newblock {\em\protect\JournalTitle{Physics Reports}} \textbf{12}, 75 -- 199
  (1974).

\bibitem{bleher1975critical}
P Bleher, YG Sinai, Critical indices for dyson's asymptotically-hierarchical
  models.
\newblock {\em\protect\JournalTitle{Communications in Mathematical Physics}}
  \textbf{45}, 247--278 (1975).

\bibitem{katzgraber:03}
H Katzgraber, AP Young, Monte carlo studies of the one-dimensional ising spin
  glass with power-law interactions.
\newblock {\em\protect\JournalTitle{Phys. Rev. B}} \textbf{67}, 134410 (2003).

\bibitem{leuzzi:08}
L Leuzzi, G Parisi, F Ricci-Tersenghi, JJ Ruiz-Lorenzo, Dilute one-dimensional
  spin glasses with power law decaying interactions.
\newblock {\em\protect\JournalTitle{Phys. Rev. Lett.}} \textbf{101}, 107203
  (2008).

\bibitem{leuzzi:09}
L Leuzzi, G Parisi, F Ricci-Tersenghi, JJ Ruiz-Lorenzo, Ising spin-glass
  transition in a magnetic field outside the limit of validity of mean-field
  theory.
\newblock {\em\protect\JournalTitle{Phys. Rev. Lett.}} \textbf{103}, 267201
  (2009).

\bibitem{jensen:21}
S Jensen, N Read, AP Young, Nontrivial maturation metastate-average state in a
  one-dimensional long-range ising spin glass: Above and below the upper
  critical range.
\newblock {\em\protect\JournalTitle{Phys. Rev. E}} \textbf{104}, 034105 (2021).

\bibitem{franz1992ultrametricity}
S Franz, G Parisi, MA Virasoro, Ultrametricity in an inhomogeneous simplest
  spin glass model.
\newblock {\em\protect\JournalTitle{Europhysics Letters}} \textbf{17}, 5
  (1992).

\bibitem{panchenko2015free}
D Panchenko, The free energy in a multi-species sherrington--kirkpatrick model.
\newblock {\em\protect\JournalTitle{The Annals of Probability}} \textbf{43},
  3494–--3513 (2015).

\bibitem{castillo:02}
HE Castillo, C Chamon, LF Cugliandolo, MP Kennett, Heterogeneous aging in spin
  glasses.
\newblock {\em\protect\JournalTitle{Phys. Rev. Lett.}} \textbf{88}, 237201
  (2002).

\bibitem{parisi2003local}
G Parisi, Local overlaps, heterogeneities and the local fluctuation dissipation
  relations.
\newblock {\em\protect\JournalTitle{Journal of Physics A: Mathematical and
  General}} \textbf{36}, 10773--10789 (2003).

\bibitem{kurchan2023time}
J Kurchan, Time-reparametrization invariances, multithermalization and the
  parisi scheme.
\newblock {\em\protect\JournalTitle{SciPost Physics Core}} \textbf{6}, 001
  (2023).

\bibitem{crisanti2023dynamical}
A Crisanti, S Franz, J Kurchan, A Maiorano, Dynamical mean-field theory and the
  aging dynamics in {\em Spin Glass Theory and Far Beyond: Replica Symmetry
  Breaking After 40 Years}.
\newblock (World Scientific), pp. 157--186 (2023).

\bibitem{bolthausen:98}
E Bolthausen, AS Sznitman, On {R}uelle's probability cascades and an abstract
  cavity method.
\newblock {\em\protect\JournalTitle{Comm. Math. Phys.}} \textbf{197}, 247--276
  (1998).

\bibitem{boettcher2005extremal}
S Boettcher, Extremal optimization for sherrington-kirkpatrick spin glasses.
\newblock {\em\protect\JournalTitle{The European Physical Journal B-Condensed
  Matter and Complex Systems}} \textbf{46}, 501--505 (2005).

\bibitem{palassini2008ground}
M Palassini, Ground-state energy fluctuations in the sherrington--kirkpatrick
  model.
\newblock {\em\protect\JournalTitle{Journal of Statistical Mechanics: Theory
  and Experiment}} \textbf{2008}, P10005 (2008).

\bibitem{boettcher2010simulations}
S Boettcher, Simulations of ground state fluctuations in mean-field ising spin
  glasses.
\newblock {\em\protect\JournalTitle{Journal of Statistical Mechanics: Theory
  and Experiment}} \textbf{2010}, P07002 (2010).

\bibitem{billoire2006numerical}
A Billoire, Numerical estimate of the finite-size corrections to the free
  energy of the sherrington-kirkpatrick model using guerra-toninelli
  interpolation.
\newblock {\em\protect\JournalTitle{Physical Review B}} \textbf{73}, 132201
  (2006).

\bibitem{aspelmeier2008finite}
T Aspelmeier, A Billoire, E Marinari, MA Moore, Finite-size corrections in the
  sherrington--kirkpatrick model.
\newblock {\em\protect\JournalTitle{Journal of Physics A: Mathematical and
  Theoretical}} \textbf{41}, 324008 (2008).

\bibitem{parisi2019study}
G Parisi, L Sarra, L Talamanca, Study of longitudinal fluctuations of the
  sherrington--kirkpatrick model.
\newblock {\em\protect\JournalTitle{Journal of Statistical Mechanics: Theory
  and Experiment}} \textbf{2019}, 033302 (2019).

\bibitem{guerra:02}
F Guerra, FL Toninelli, The thermodynamic limit in mean field spin glass
  models.
\newblock {\em\protect\JournalTitle{Communications in Mathematical Physics}}
  \textbf{230}, 71--79 (2002).

\bibitem{panchenko2015hierarchical}
D Panchenko, Hierarchical exchangeability of pure states in mean field spin
  glass models.
\newblock {\em\protect\JournalTitle{Probability Theory and Related Fields}}
  \textbf{161}, 619--650 (2015).

\bibitem{panchenko2016structure}
D Panchenko, Structure of finite-rsb asymptotic gibbs measures in the diluted
  spin glass models.
\newblock {\em\protect\JournalTitle{Journal of statistical physics}}
  \textbf{162}, 1--42 (2016).

\bibitem{contucci:09}
P Contucci, C Giardin{\`a}, C Giberti, G Parisi, C Vernia, Structure of
  correlations in three dimensional spin glasses.
\newblock {\em\protect\JournalTitle{Phys. Rev. Lett}} \textbf{103}, 017201
  (2009).

\bibitem{janus:10}
R Alvarez~Ba{\~n}os, et~al., Nature of the spin-glass phase at experimental
  length scales.
\newblock {\em\protect\JournalTitle{J. Stat. Mech.}} \textbf{2010}, P06026
  (2010).

\bibitem{kotliar:85}
G Kotliar, PW Anderson, DL Stein, One-dimensional spin-glass model with
  long-range random interactions.
\newblock {\em\protect\JournalTitle{Phys. Rev. B}} \textbf{27}, 602 (1983).

\bibitem{banos:12}
RA Ba\~nos, LA Fernandez, V Martin-Mayor, AP Young, Correspondence between
  long-range and short-range spin glasses.
\newblock {\em\protect\JournalTitle{Phys. Rev. B}} \textbf{86}, 134416 (2012).

\bibitem{angelini2014relations}
MC Angelini, G Parisi, F Ricci-Tersenghi, Relations between short-range and
  long-range ising models.
\newblock {\em\protect\JournalTitle{Physical Review E}} \textbf{89}, 062120
  (2014).

\bibitem{de1985gaussian}
C De~Dominicis, I Kondor, Gaussian propagators for the ising spin glass below
  tc.
\newblock {\em\protect\JournalTitle{Journal de Physique Lettres}} \textbf{46},
  1037--1043 (1985).

\bibitem{parisi1992finite}
G Parisi, P Biscari, Finite-volume corrections to the mean-field solution of
  the sk model.
\newblock {\em\protect\JournalTitle{Journal of Physics A: Mathematical and
  General}} \textbf{25}, 4787 (1992).

\bibitem{aguilar2024small}
M Aguilar-Janita, et~al., Small field chaos in spin glasses: Universal
  predictions from the ultrametric tree and comparison with numerical
  simulations.
\newblock {\em\protect\JournalTitle{Proceedings of the National Academy of
  Sciences}} \textbf{121}, e2404973121 (2024).

\bibitem{brunetti1992asymmetric}
R Brunetti, G Parisi, F Ritort, Asymmetric little spin-glass model.
\newblock {\em\protect\JournalTitle{Physical Review B}} \textbf{46}, 5339
  (1992).

\bibitem{franz2003replicaI}
S Franz, M Leone, Replica bounds for optimization problems and diluted spin
  systems.
\newblock {\em\protect\JournalTitle{Journal of Statistical Physics}}
  \textbf{111}, 535--564 (2003).

\bibitem{franz2003replicaII}
S Franz, M Leone, FL Toninelli, Replica bounds for diluted non-poissonian spin
  systems.
\newblock {\em\protect\JournalTitle{Journal of Physics A: Mathematical and
  General}} \textbf{36}, 10967 (2003).

\bibitem{1742-5468-2014-3-P03019}
M Picco, N Sourlas, On the phase transition of the 3d random field ising model.
\newblock {\em\protect\JournalTitle{Journal of Statistical Mechanics: Theory
  and Experiment}} \textbf{2014}, P03019 (2014).

\bibitem{mezard:84b}
M M{\'e}zard, G Parisi, N Sourlas, G Toulouse, M Virasoro, Replica symmetry
  breaking and the nature of the spin glass phase.
\newblock {\em\protect\JournalTitle{J. Phys. France}} \textbf{45}, 843--854
  (1984).

\end{thebibliography}
\clearpage

\begin{center}
\textbf{\huge Supplementary Information}

\textbf{
\LARGE 
Some technical details
}    
\end{center}

\section*{Finite volume corrections for the $O(n)$ ferromagnetic model}

In the high-temperature region, the Hessian around the saddle point is positive definite, and the fluctuations are Gaussian. 
For example, in the ferromagnetic model, we have the following representation for the free energy 
\begin{multline}
    -\beta f(\beta,N)=\frac1N
    \log\left(\int
    \frac{d\vec{m}}{({2\pi/(\beta N)})^{n/2}} e^{-\beta N F(m)}\right)\approx\\
    \approx -\beta F(m^*)+ \frac{1}{2N} {\rm Tr}\; \log({\cal H})
\end{multline}
where the second term is produced by the Gaussian fluctuations around the saddle point $m^*$ and $\cal H $
is the Hessian matrix describing such small fluctuations. 
In the low-temperature region, the $1/N$ scaling is unchanged even in the presence of symmetry-breaking zero modes in the Hessian, yielding only logarithmic corrections.
The degenerate saddle points are simply taken into account by integrating over the degrees of freedom of the symmetry group, leading to
\begin{multline}
   -\beta f(\beta,N) = -\beta F(m^*)- \frac{1}{2N} {\rm Tr}\;\log({\cal H_\perp})\\ 
   +\frac{1}{N}\log(\Omega_{n}(m^*)) -\frac{n-1}{2 N}\log(2\pi\beta N)
\end{multline}
where ${\cal H_\perp}$ is the transverse Hessian, $\Omega_{n}(m^*)$ is the surface of the $n$-dimensional sphere of radius $m^*$ and the $\log N$ terms comes from the $n-1$ zero modes.

\section*{Correlation function in the hierarchical model}

Here, we show the connection (in the ordered phase, where mode locking takes place) between the connected correlation function measure in the hierarchical model and the function $D(t,N)$. For the reader's convenience, we recall some properties of the hierarchical model and of its correlation.

In the hierarchical model, we can introduce a sort of dyadic distance $d$ between two different points $i$ and $k$ in such a way that for a system of size  $N=2^L$ ($L$ being the number of blocks), there are exactly $2^{r}$ points at distance $d=2^r$ ($r=1,L-1$). Sites that are in the same smallest block are at a distance of 1, sites that are in the same smallest block but not in the first are at a distance of 2, and so on.

If we consider a simple scalar quantity $\sigma_i$, with a standard ferromagnetic Hierarchical Hamiltonian, we have that 
\begin{equation}
    \langle \sigma_i \sigma_k \rangle= G(d_{i,k})\,.
\end{equation}
Therefore, the correlation function can take at most $L$ distinct values (including the case $i=k$, which corresponds to a distance of 0).

We define
\begin{equation}
    \Sigma_L=\sum_{i=1}^{N/2}\sigma_i\,, \quad
    \Sigma_R=\sum_{i=N/2+1}^N\sigma_i\,,
\end{equation}
and the spins in the two $\Sigma$'s stay at distance $2^{L-1}$ from each other.

We now have
\begin{equation}
    \langle (\Sigma_L-\Sigma_R)^2\rangle =
    \langle (\Sigma_L^2+\Sigma_R^2- 2\; \Sigma_L \,\Sigma_R\rangle\,,
\end{equation}
and
\begin{equation}
    \langle \Sigma_L\Sigma_R\rangle =2^{2(L-1)}G(2^{L-1})\,,
\end{equation}
while 
\begin{equation}
 \langle \Sigma_L^2\rangle=\langle\Sigma_R^2\rangle =
 2^{L-1}\left( \langle\sigma^2_i\rangle+\sum_{r=1}^{L-2}2^{r} G(2^r) \right )\,,
\end{equation}
where the factor $2^r$ comes from the number of points at distance $r$ and $\langle\sigma^2_i\rangle$ is the contribution of the points at distance 0.
 
We can now neglect the term $\langle\sigma^2_i\rangle=1$ in the region where the correlation length is large. If we put everything together, we get
\begin{eqnarray}
 && \Delta(L)\equiv 2^{-2(L-1)-1} \langle (\Sigma_L-\Sigma_R)^2\rangle=\nonumber\\
  &&\sum_{r=1}^{L-2}2^{-(L-1-r)}G(2^r) - G(2^{L-1}) =\nonumber \\
   && \sum_{s=1}^{L-2}2^{-s}G(2^{L-1-s}) - G(2^{L-1})\approx\nonumber\\
   && \sum_{s=1}^{L-2}2^{-s}\left(G(2^{L-1-s}) - G(2^{L-1})\right)
\end{eqnarray}
If the correlation function is a decreasing function of the distance, the previous expression is automatically positive, as it should be. A nice property of this quantity is that it depends only on the difference of correlation functions, so that it does not matter if we use the connected or the disconnected correlation functions.

Let us make clear why $\Delta(L)$ is a proxy for the \emph{connected} correlation function $G_c(d)$ at large distances. We find that $\Delta(L) \approx G_c(d(L))$ when the connected correlations go to zero as a power of the distance (we have introduced the notation $d(L) \equiv 2^{L-1}$)
\begin{itemize}
     \item First, we consider the situation where the correlation has a power behavior,  $G(d)\propto d^{-\alpha}$ with $\alpha<2$. In this case, $\langle \Sigma_L^2\rangle$ diverges as $d(L)^{2-\alpha}$ and $\Delta(L) \propto d(L)^{-\alpha} $.
     \item Second, in case $G(d)=A+G_c(d)$, with the connected correlation function $G_c(d)$  decaying as a power-law $G_c(d)\propto d^{-\alpha}$ with $\alpha<2$, we still find that $\Delta(L) \propto G_c(d(L)) $. Indeed $G(2^{L-1-s}) - G(2^{L-1}) = G_c(2^{L-1-s}) - G_c(2^{L-1})$.
\end{itemize}

Let us go back to spin glasses: we define
\begin{equation}
    q_L=\frac 2 N \sum_{i=1}^{N/2}q_i\,, \quad
    q_R=\frac 2 N \sum_{i=N/2+1}^N q_i\,,
\end{equation}
The previous discussion shows that in the hierarchical model 
\begin{equation}
    \mathbb{E}(\langle (q_L-q_R)^2\rangle) \,
    \end{equation}
is proportional to the connected correlation function, as discussed in the text.

Let us now condition the global overlap $q=(q_L+q_R)/2$.
For the connected correlation we have
\begin{equation}
    G_c(2^{L-1}|q)=\mathbb{E}(\langle q_L q_R|q\rangle_c)= \mathbb{E}(\langle q_L q_R|q\rangle)-q^2\,.
\end{equation}
since $\mathbb{E}(\langle q_L|q\rangle)=\mathbb{E}(\langle q_R|q\rangle)=q$ and $q_R=2q-q_L$.
On the other hand, $D(t,N|q)$ is defined as $D(t,N|q)=\mathbb{E}(\langle s^2|q\rangle_{t,N})$ with $s=q_L-q_R$ and we can rewrite
\begin{multline}
D(t,N|q)=\mathbb{E}(\langle (q_L-q_R)^2|q\rangle) = 4\, (q^2 - \mathbb{E}(\langle q_L q_R|q\rangle) 
\end{multline}
From the above expressions, it is clear that $D(t,N)$ is proportional to the connected correlation function at a distance $x-y \sim N$.

\section*{A more detailed scaling analysis}

Let us reconsider the scaling analysis of $\mathbb{E}\langle (q^L-q^R)^2\rangle$,
imagining a more general situation where we have the following asymptotic regimes 
of the functions ${\cal D}(z)$ for $z\to\infty$ and 
$D(t,N)$ for $t\to 0$
\begin{equation}
    {\cal D}(z)\approx 1/z^a\qquad D(t,N) \approx t^{-c}/N =z^{-c} N^{c-1}
\end{equation}
The two functions are of the same order when 
\begin{equation}
 z\approx   z^*(N)=N^{(c-1)/(c-a)} \,.
\end{equation}
An appropriate scaling function should describe the cross-over.
Let us define  $\hat D(z,N) \equiv D(z/N,N)  =1/z^a$ for $z\ll Z(N)$ and $\hat D(z,N)  =z^{-c} N^{c-1}$ for $z\gg Z(N)$.
We have to evaluate the correction
\begin{equation}
   N^{-1} \int_0^N dz \hat D(z,N)\,.
\end{equation}
If we use $z^*(N)$ to split the integral into two contributions, and suppose that the cross-over region gives a sub-dominant contribution.
We get two terms, the first given by 
\begin{multline}
N^{-1}\int_0^{z^*(N)}dz\; \hat{D}(z,N) \propto  (1-a)^{-1}N^{-1} z^*(N)^{1-a} \\
\propto N^{(1-a)(c-1)/(c-a)-1}\,.
\end{multline}
The second term is given by
\begin{align}
   N^{-1}\int_{z^*(N)}^\infty dz\; \hat{D}(z,N)
\end{align}
and it is asymptotically proportional to the first one.

Let us consider the exponent in the first term. As we have seen, in the SK model $c=3/2$ so we get $N^{-\omega}$  with 
\begin{equation}
\omega= 1-\frac{1-a}{3-2\,a}.
\end{equation}
For $a=1/2$, as suggested by our semi-analytic results, we recover $\omega=3/4$, while a correction $a=1/2+\epsilon$ would imply $\omega=3/4+\epsilon/8$ for small $\epsilon$.

\begin{figure}[t]
\centering
\includegraphics[width=\linewidth]{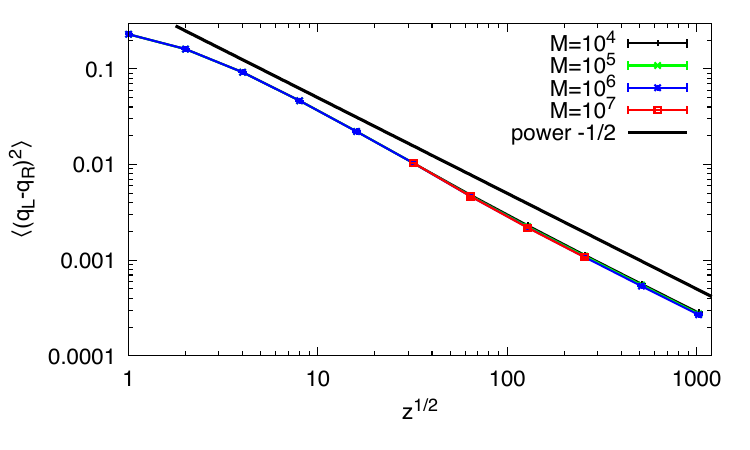}
 \caption{The semi-analytic predictions for ${\cal D}(z) $ versus $z^{1/2}$ $M=10^4,10^5,10^6,10^7$. The black line is proportional to $z^{1/2}$}
 \label{fig:A1}
\end{figure}

If we fit the data for $N=10^6$ with a simple power law for $ z\ge 32$, we obtain a poor fit, with an exponent  $a=0.54\pm 0.01$. If we remove the point at $z=32$ from the fit and keep only the points at $z\ge 64$, we obtain a better fit, with $a=0.52\pm 0.005$. We think that we can safely conclude that $a$ is close to $1/2$, and it is certainly different from zero.

\begin{figure}[t]
\centering
\includegraphics[width=\linewidth]{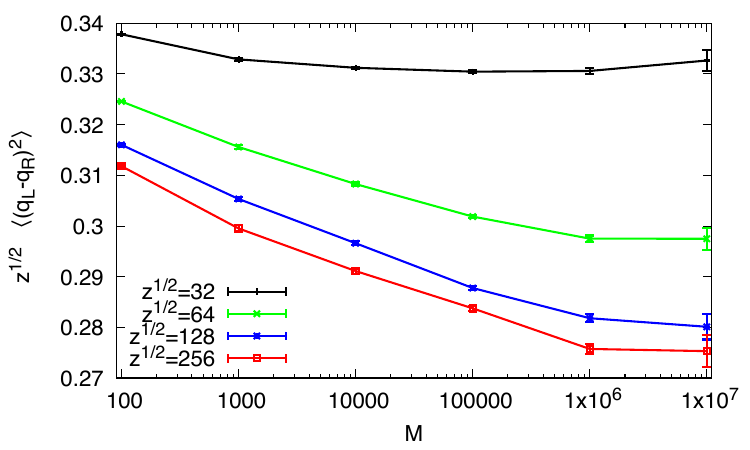}
 \caption{The semi-analytic predictions for $z^{1/2}{\cal D}(z) $ versus $M$ for $z^{1/2}=32,64,128,256$.} 
 \label{fig:A2}
\end{figure}

For large values of $z$ it is not easy to do a precise extrapolation to $M=\infty$, however, the $M$ dependence of ${\cal D}(z)$ is weak, as clearly shown in Fig.~\ref{fig:A2}: the variation from $M=10^6$ to $M=10^7$ is less than 1\%.

The exponent $\omega=2/3$ quoted in the previous literature would require $a=0$, which, as we discussed in the main text, is incompatible with the fact that locking occurs on a finite $z$ scale and ${\cal D}(z)$ vanishes for $z\to\infty$ as clearly shown in Fig.~\ref{fig:A1}. Notice that $1/N$ terms pop out both in the region of $z$ finite and the region of $t$ finite. Such a term has to be added in the fit, and its inclusion strongly increases the error in the value of the exponent.

\section*{The full distribution function}

We have seen that, when we have two subsystems, we can measure the corresponding overlaps $q_L$ and $q_R$, and it is convenient to introduce the two linear combinations $q=(q_L+q_R)/2$ and $s=q_L-q_R$. We can thus define the probability distribution of $|q|$, and $s$ that we call ${\cal P}(q,s;N)$.
Obviously, this function also depends on $z$ or $t\equiv z/N$, but we have not indicated this dependence to lighten the notation.

We can restrict the analysis to the $|q|$ because we know that in the absence of a magnetic field, the probability distribution of $q$ is symmetric around the origin. Up to now, we have studied the quantity
\begin{equation}
    \int dq\, d s\; {\cal P}(q,s;N)\, s^2 \,.
\end{equation}
There are other quantities that can be studied, notably
\begin{equation*}
  {\cal P}(q;N)\equiv  \int d s \;{\cal P}(q,s) \quad { D}(t,N|q)\equiv\frac{\int d s \;{\cal P}(q,s;N)s^2}{\int d s \;{\cal P}(q,s;N)}\,.
\end{equation*}
The last quantity, $D(t,N|q)$, is crucial to compute the correlation functions and to recover the wonderful results of Ref.~\cite{jensen:21} on the correlation as discussed in the main text.

In the limit where $N\to \infty$ at fixed t, $P(q;N) \to P(q)$ where the $P(q)$ is the usual probability density function
\begin{equation}
    P(q)=(1-x_M)\delta(q-q_\text{\tiny EA})+p(q)
\end{equation}
where $p(q)$ is a smooth function.
The simplest hypothesis is that for this function, the crossover function is trivial, and that 
\begin{equation}
    \lim_{N\to\infty, z \text{ fixed}} {\cal P}(q,N)\equiv P(q,z) \quad \lim_{z\to\infty} P(q,z)= P(q)
\end{equation}

Let us examine whether this hypothesis is supported by the semi-analytic analysis. We readily find at $z=0$ that 
\begin{equation}
    P(q,0)=\frac{(1-x_M)^2}{2}\left(\delta(q)+\delta (q-q_\text{\tiny EA})\right)+p(q) \,,
\end{equation}
that is quite different from the asymptotic form that has only one delta function. 

\begin{figure}[t]
\centering
\includegraphics[width=\linewidth]{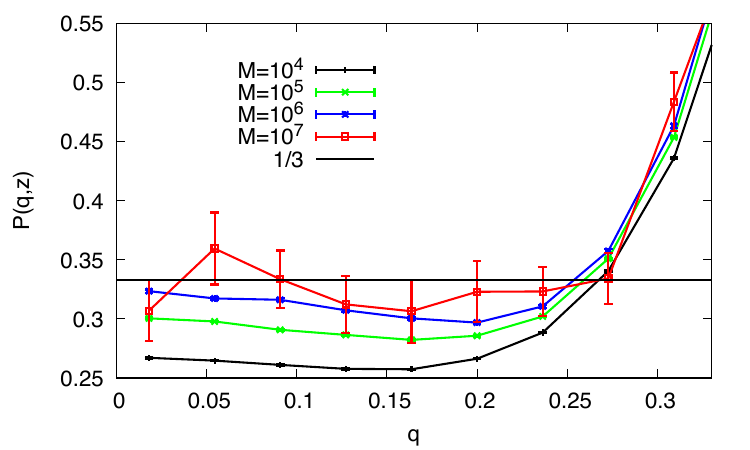}
 \caption{The semi-analytic results for $P(q,z)$ at $z^{1/2}=64$ for $M=10^4,10^5,10^6,10^7$. The horizontal line is at $1/3$, i.e., the supposed asymptotic value for $P(q)$.}
 \label{fig:A3}
\end{figure}

The constrained averages at fixed $q$ unfortunately converge more slowly than the unconstrained averages,which is the main reason that has pushed us to do simulations with such a large value of $M$ (in the Ref.~\cite{1742-5468-2014-3-P03019} the maximum value was $M=10^5$). We will now present only a few examples to support our claims. In Fig.~\ref{fig:A3}, we show the semi-analytic results for $P(q,z)$ at $z^{1/2}=64$, the largest value of $z$ for which we achieve convergence. On the right, for $q>0.3$, we see that the function increases; this is due to the smearing of the peak near $q_\text{\tiny EA}=0.4$.

\begin{figure}[t]
\centering
\includegraphics[width=\linewidth]{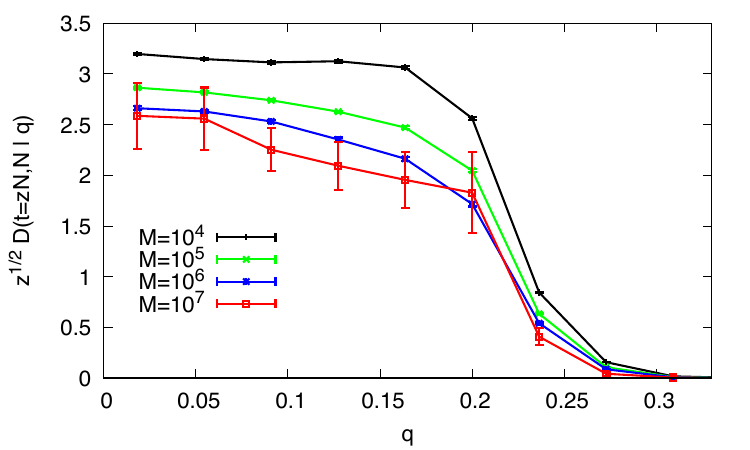}
 \caption{The semi-analytic results for $z^{1/2}D(t=zN,N|q)$ at $z^{1/2}=64$ for $M=10^4,10^5,10^6,10^7$.}
 \label{fig:A4}
\end{figure}

Having assured that at $z^{1/2}=64$ the function $P(q,z)$ is not too far from its asymptotic limit, we can now look at the function $D(q,z)$ at the same value of $z$.  This function is depicted in Fig.~\ref{fig:A4}. It is quite clear that this function does not build a singularity at $q=0$ (neither wants to go to zero). Therefore, our hypothesis in the main text on the regularity of the function $D(q,z)$ near $q=0 $ is fully justified.

\section*{The generation of the correlated random noise $\Delta$}

We observe that in the presence of $\mathbb{Z}_2$ spin inversion symmetry, each state has a double where all the spins are inverted, and the overlap probability distributions must be symmetric around zero. 
The tree of states in spin glasses has been described in detail \cite{mezard:84b}.
In spin glasses, the set of states of a given instance is described by the weights $w_\alpha$ of the states ($\sum_\alpha w_\alpha=1$) and by the overlaps $q_{\alpha,\gamma}$, with $q_{\alpha,\alpha}=q_\text{\tiny EA}$. 
As we will see later, it is often convenient to group pairs of states related by symmetry in such a way that only positive overlaps need to be considered. The property of ultrametricity implies that the states can be viewed as the leaves of a tree, which exhibits peculiar statistics. Overlaps are defined as the minimum level that must be crossed to transition from one level to another.
 
\begin{figure}[t]
\centering
\includegraphics[width=0.95\linewidth]{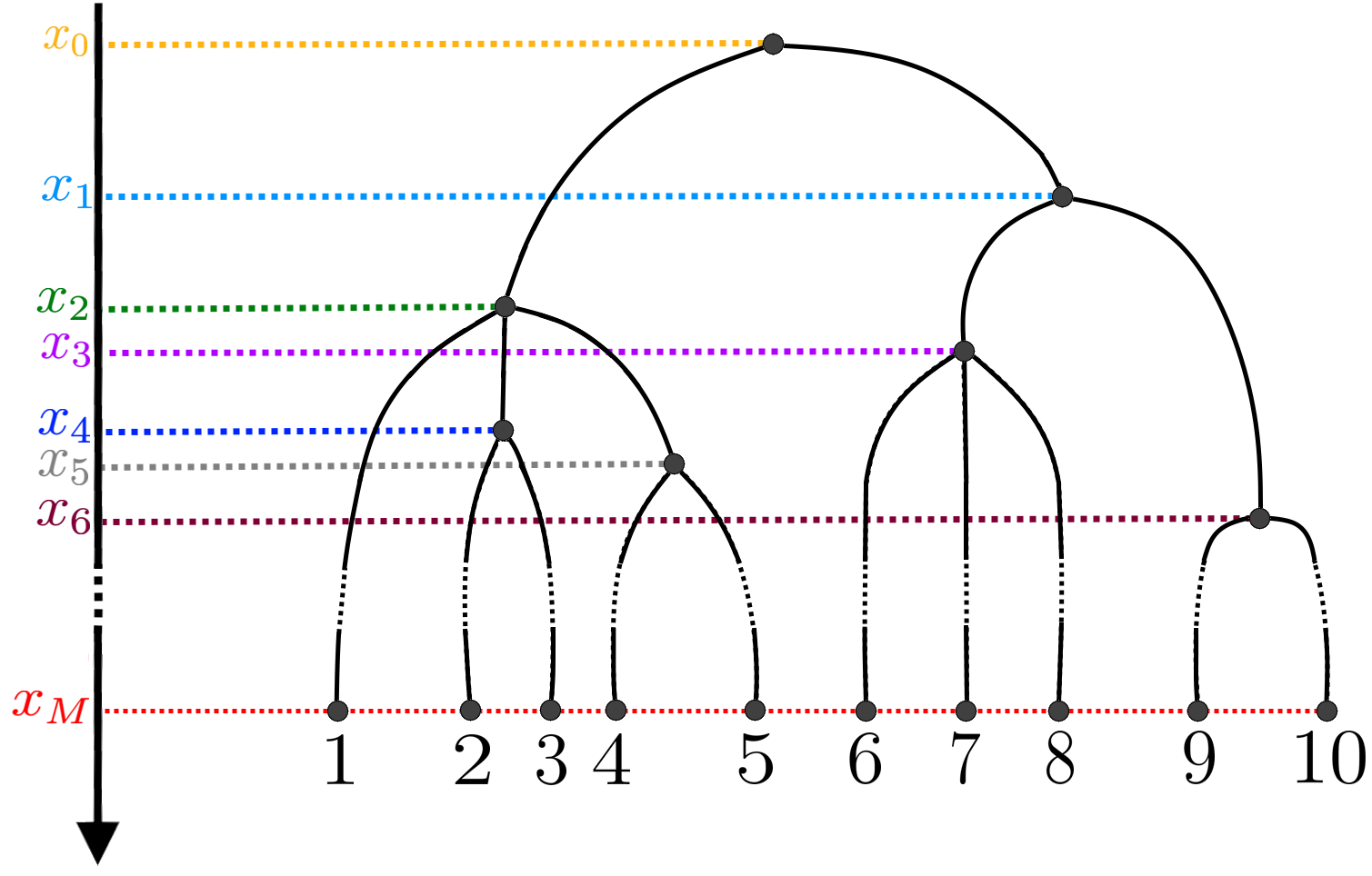}
\caption{A tree with ten leaves. The overlap of leave (1) with the leaves ($2\cdots5$) is $q(x_2)$ and with the leaves ($6\cdots 10$) is $q(x_0)$; the overlaps of leave (2) with the other leaves are the same as leave (1), with the only difference that the overlap with the leave (3) is now $q(x_4)$ (taken from \cite{aguilar2024small}.}
\label{fig:tree}
\end{figure}

The tree has infinitely many leaves, but with a cutoff on the smallest weights, it can be approximated by a tree with a finite number of leaves $M$. 
The generation of a tree has been explained in detail in Ref.~\cite{aguilar2024small} and will not be reproduced here; we only reproduce from that paper a figure depicting a tree with ten leaves (fig \ref{fig:tree}). 
We use their approach to extract, with the correct probability, a tree with $M$ leaves: we must determine the weights of the leaves and the mutual overlaps (which are in one-to-one correspondence with the levels of the branching points). 
In our case, we have to compute 
\begin{equation}
    I(z)=\mathbb{E}\log\left(\sum w_\alpha^L \,w_\gamma^R \,\exp( -\beta \sqrt{z} \Delta_{\alpha,\gamma}) \right)\,,
\end{equation}
where the  $M$ weights $ w_\alpha^L $ and $ w_\gamma^R $ and the overlap $q^L_{\alpha,\alpha'}$ and $q^R_{\gamma,\gamma'}$ of the left and right trees have been generated numerically following the procedure described in \cite{aguilar2024small}.  The energies $\Delta_{\alpha,\gamma}$ associated with a pair of leaves $\alpha,\gamma$ (which we call bileaves) are zero-mean Gaussian variables with correlations 
\begin{align}
\mathbb{E}(\Delta_{\alpha,\gamma}\Delta_{\alpha',\gamma'})=q^L_{\alpha,\alpha'} q^R_{\gamma,\gamma'}\,,
\label{eq:corr}
\end{align}
with $q_{\alpha,\alpha}=q_\text{\tiny EA}$.
It is easy to see, using the properties of Gaussian variables, that $I(z)$ verifies
\begin{eqnarray}
    \frac{dI(z)}{dz}=\frac{\beta^2}{2}\left(q_\text{\tiny EA}^2-\langle q_L q_R\rangle\right)\,. 
\end{eqnarray}
from which we can reconstruct ${\cal D}(z)$.

The code we used is deposited in a repository (https://github.com/giorgioparisi-cmyk/SyncroGlass.git). The only tricky part in the computation is generating the $\Delta$'s with the correct correlations. In Ref.~\cite{1742-5468-2014-3-P03019}, we already explained how to create correlated variables with prescribed variance on a single tree.  Similarly, we can consider elements of the direct product of the nodes of each tree; these elements will be called binodes.
An efficient way of generating Gaussian variables $\Delta_{\alpha\gamma}$ with correlations in Eq.~\ref{eq:corr} can be achieved considering the paths of nodes leading from the root of the trees to the leaves $\alpha$ and $\gamma$ on the coupled trees, say $\alpha\to \{\alpha_0\equiv0,\alpha_1,\ldots,\alpha_k\equiv\alpha\}$ and $\gamma\to \{\gamma_0\equiv0,\gamma_1,\ldots,\gamma_\ell\equiv\gamma\}$. 
One can associate to each binode ${\alpha_i,\gamma_j}$ an energy $r(\alpha_i,\gamma_j)$ and define 
\begin{eqnarray}
\Delta_{\alpha\gamma}=\sum_{i=0}^{k}\sum_{j=0}^{\ell}r(\alpha_i,\gamma_j). 
\end{eqnarray}
It is easy to see that if  we choose the variables $r$ as independent Gaussian variables with correlations that depend only on the levels 
\begin{eqnarray}
    V(\alpha_i,\gamma_j) \equiv \mathbb{E}(r(\alpha_i,\gamma_j)^2)=(q_i-q_{i-1})(q_j-q_{j-1})
    \end{eqnarray}
(with the convention that 
$q_{-1}=0$) we produce $\Delta$ variables with the right statistics. In this way, the sums of energies from the biroots to a binode need to be computed only once.

The construction is complex. Let us see in detail how it works. 
We consider first the case where both trees have only the root and leaves;  the nodes of one tree will be the root (0), and the leaves labeled by $\alpha$; the nodes of the other tree will be the root (0), and the leaves labeled by $\gamma$. Let us suppose that the roots have an overlap $q_0$ and the leaves $q_1$ (the same for both trees, just for simplicity). The binodes will be
\begin{align*}
    0,&0 \\ 
    \alpha,0 \quad&\quad 0,\gamma \\ 
    \alpha,&\gamma
\end{align*}
We set 
\begin{equation}
    \Delta_{\alpha\gamma}=r(0,0)+ r(\alpha,0)+ r(0,\gamma)+r(\alpha,\gamma)
\end{equation}
where the $r$'s are independent random Gaussian variables with zero mean and variances given by
\begin{align}
    &V(0,0)=q_0^2\,,\nonumber\\
    &V(\alpha,0)=V(0,\gamma)=q_0(q_1-q_0)\,,\\
    &V(\alpha,\gamma)=(q_1-q_0)^2\,.\nonumber
\end{align}
In the case where each tree has a root ($0$), clusters labeled by $\alpha_1$ in the first tree (resp.\ $\gamma_1$ in the second tree), and leaves labeled by $\alpha_2$ in the first tree (resp.\ $\gamma_2$ in the second tree),the binodes are
\begin{align*}
0,&0 \\
\alpha_1,0 \quad&\quad 0,\gamma_1 \\
\alpha_2,0 \qquad \alpha_1,&\gamma_1 \qquad 0,\gamma_2\\
\alpha_2,\gamma_1 \quad&\quad \alpha_1,\gamma_2 \\
\alpha_2,&\gamma_2
\end{align*}
We thus have 
\begin{multline}
    \Delta_{\alpha_2,\gamma_2} = r(0,0) + r(\alpha_1,0) + r(0,\gamma_1) +\\
    r(\alpha_2,0) + r(\alpha_1,\gamma_1) + r(0,\gamma_2) +\\
    r(\alpha_2,\gamma_1) + r(\alpha_1,\gamma_2) + r(\alpha_2,\gamma_2)
\end{multline}
where the variances are given by
\begin{align}
   & V(0,0)=q_0^2\,,\nonumber\\
   & V(\alpha_1,0)=V(0,\gamma_1)=q_0(q_1-q_0)\,,\nonumber\\
   & V(\alpha_1,\gamma_1)=(q_1-q_0)^2\,,\nonumber\\
   & V(\alpha_2,0)=V(0,\gamma_2)=q_0(q_2-q_1)\,,\nonumber\\
   & V(\alpha_2,\gamma_1)=V(\alpha_1,\gamma_2)=(q_1-q_0)(q_2-q_1)\,,\nonumber\\
   & V(\alpha_2,\gamma_2)=(q_2-q_1)^2\,.
\end{align}
The construction can be easily generalized to many levels, finding that 
\begin{equation}
    \Delta_{\alpha,\gamma}=\sum_{\alpha'\in A_1(\alpha);\gamma'\in A_2(\gamma)} r(\alpha',\gamma') \,,
\end{equation}
where $A_1(\alpha)$ is the set of the ancestors of $\alpha$ on the first tree and $A_2(\gamma)$ is the set of the ancestors of $\gamma$ on the second tree. In other words, the sum is over all the biancestors of the bileaves. The quantities $r(\alpha,\gamma)$ are random independent variables with variance
\begin{equation}
    V(\alpha,\gamma)=\big(q(\alpha)-q(p_1(\alpha))\big)\big(q(\gamma)-q(p_2(\gamma))\big)\,,
\end{equation}
where one denotes $p_1(\alpha)$ and $p_2(\gamma)$ the parents of the nodes $\alpha$ and $\gamma$ on their respective tree, and $q(\alpha)$ denotes the overlap at the level of the node $\alpha$. 

This formula is sufficient to produce the final results, but when applied to all binodes, some parts of the computation are duplicated. Fast computation can be achieved if we can associate energy not only with the bileaves but also with the binodes. In this way, one finds a nifty recursion formula:
\begin{eqnarray}
    \Delta_{\alpha,\gamma}= \Delta_{p_1(\alpha),\gamma} + \Delta_{\alpha,p_2(\gamma)} - \Delta_{p_1(\alpha),p_2(\gamma)} + r_{\alpha,\gamma} \,,   
\end{eqnarray}
where $r_{\alpha,\gamma}$ is the same as before; the subtraction of the term $\Delta_{p_1(\alpha),p_2(\gamma)}$ corrects for the double counting present in the sum of the first two terms. We can use this recursive equation to compute the energies, starting from the root, thereby avoiding redundant computations.

\section*{Numerical implementation}

We want to compute $I(z)$ when $M=O(10^7)$ and the number of levels is arbitrary, so some care is needed for writing an efficient code.

We represent each tree as a list of nodes, including the leaves. Each node has an ID, the value $x$ of its depth, and the ID of the parent. The nodes are ordered from the root to the leaves by increasing value $x$.

In the case of the direct product of two trees, each binode has an ID, the ID of the two nodes in the two original trees, and the ID of the two binodes that are immediately higher in the hierarchy (to the left and to the right), i.e., the two biparents. In this way, we can sort the binodes in such a way that the two biparents of the binode appear before the binode in the list. In this way, we can traverse the whole tree starting from the biroot in a way similar to the one used in Ref.~\cite{aguilar2024small}.

The value of the weights of the bileaves at $z=0$ is $w_\alpha^L w_\gamma^R$, where the $w$'s are typically in the range $(\epsilon,1)$ with $\epsilon =O(M^{-1/x_M})$ and we may have bileaves weights of order $\epsilon^2$. We can reduce the computational complexity by retaining only the bileaves whose weight exceeds $\epsilon$. In this way, the number of bileaves decreases from $M^2$ to $O(M\log(M)$. Everywhere, all probabilities that are smaller than $\epsilon$ can be set to zero.

\section*{Perturbative evaluation of the mode locking integral}

In setting up a perturbative expansion for ${\cal D}(z)$, we have two alternatives. Either we use the 
probabilistic setting, we exploit the statistical properties of the weights, or we use the replica approach. For simplicity, we present the direct approach here. The replica approach for computing more orders in $z$ will be presented elsewhere. We have verified that the direct approach reproduces the first orders of the perturbative expansion with high accuracy, providing a stringent test of the computation's correctness. 

In order to compute 
\begin{eqnarray}
I(z)=\mathbb{E}\left(\log\sum_{\alpha\gamma}w_\alpha^L w_\gamma ^R e^{\beta \sqrt{z}\Delta_{\alpha\beta}}\right)\,,
\end{eqnarray}
we define the average over the weights
\begin{equation}
    [A]=\sum_{\alpha\gamma} w_\alpha^L w_\gamma ^R \; A_{\alpha\gamma}
\end{equation}
and write
\begin{eqnarray}
    I(z)=\mathbb{E\log(}[e^{\beta\sqrt{z}\Delta}]).
\end{eqnarray}
We can then expand in cumulants $[\Delta^k]_c$ as
\begin{eqnarray}
I(z)=\sum_{k {\rm \; even}
}^{0,\infty}\frac{(\beta\sqrt{z})^k}{k!}
    \mathbb{E}([\Delta^k]_c).
\end{eqnarray}
Notice that thanks to the average $[\cdot]$ over Gaussian $\Delta$ only the terms with even $k$ are present in the sum. Moreover, if $\mathbb{Z}_2$ symmetry is present, only even moments appear in the cumulants.
Thanks to this expression, we can efficiently compute the first few terms. 
To the lowest order, we find 
\begin{itemize}
    \item In absence of the $\mathbb{Z}_2$ symmetry
    \begin{equation}
    {\cal D}(z)= 2\left(\langle q^2\rangle-\langle q \rangle^2\right)
    -\frac43 \beta^2 z \left(\langle q^2\rangle-\langle q \rangle^2\right)^2 + O(z^2)\nonumber
    \end{equation}
    \item In presence of the $\mathbb{Z}_2$ symmetry
    \begin{equation}
    {\cal D}(z) = 2 \langle q^2 \rangle (1-\beta^2 z \langle q^2 \rangle) + O(z^2) \nonumber
    \end{equation}
\end{itemize}
Notice that in both cases ${\cal D}(z)$ starts to decrease linearly for small $z$. 
Using this expansion, for an explicit choice of the function $q(x)$, we could go numerically to the sixth order in $z$ in the computation of ${\cal D}(z)$.
In this way, we have checked, with and without the $\mathbb{Z}_2$ symmetry, the consistency of the semi-analytical generation of spin-glass trees with the exact results of power expansion in $z$ in the case $q(x)=3\,x$ with $q_\text{\tiny EA}=0.4$ and $\beta=1.5$. 

\end{document}